\documentclass[acmsmall,screen]{acmart}

\usepackage{bm}
\def\method{Diff-POI}

\usepackage{multirow}
\usepackage{enumitem}
\usepackage{tabularx}
\newcolumntype{Y}{>{\small\centering\arraybackslash}X}
\usepackage{graphicx}
\usepackage{subcaption}
\usepackage{stfloats}

\usepackage{algorithm}
\usepackage{algorithmic}
\usepackage[algo2e,ruled,linesnumbered]{algorithm2e}

\newcommand{\R}[1]{{\color{black}{\sc} #1}}

\AtBeginDocument{%
  \providecommand\BibTeX{{%
    \normalfont B\kern-0.5em{\scshape i\kern-0.25em b}\kern-0.8em\TeX}}}

\begin{document}

\title{A Diffusion model for POI recommendation}


\author{Yifang Qin}
\orcid{0000-0002-7520-8039}
\email{qinyifang@pku.edu.cn}
\affiliation{
    \institution{School of Computer Science, National Key Laboratory for Multimedia Information Processing, Peking University, Beijing}
    \country{China}
    \postcode{100871}
}

\author{Hongjun Wu}
\orcid{0009-0008-4173-8390}
\email{dlmao3@stu.pku.edu.cn}
\affiliation{
    \institution{School of EECS, Peking University, Beijing}
    \country{China}
    \postcode{100871}
}

\author{Wei Ju}
\authornote{Corresponding authors.}
\email{juwei@pku.edu.cn}
\orcid{0000-0001-9657-951X}
\affiliation{
    \institution{School of Computer Science, National Key Laboratory for Multimedia Information Processing, Peking University, Beijing}
    \country{China}
    \postcode{100871}
}

\author{Xiao Luo}
\email{xiaoluo@cs.ucla.edu}
\orcid{0000-0002-7987-3714}
\affiliation{
    \institution{Department of Computer Science, University of California, Los Angeles}
    \country{USA}
    \postcode{90095}
}

\author{Ming Zhang}
\authornotemark[1]
\email{mzhang\_cs@pku.edu.cn}
\orcid{0000-0002-9809-3430}
\affiliation{
    \institution{School of Computer Science, National Key Laboratory for Multimedia Information Processing, Peking University, Beijing}
    \country{China}
    \postcode{100871}
}
\renewcommand{\shortauthors}{Qin, et al.}



\received{10 April 2023}
\received[revised]{16 June 2023}
\received[accepted]{29 August 2023}

\begin{abstract}
    Next Point-of-Interest (POI) recommendation is a critical task in location-based services that aim to provide personalized suggestions for the user's next destination. Previous works on POI recommendation have \R{laid focus} on modeling the user's spatial preference. However, existing works that leverage spatial information are only based on the aggregation of users' previous visited positions, which discourages the model from recommending POIs in novel areas. This trait of position-based methods will harm the model's performance in many situations. Additionally, incorporating sequential information into the user's spatial preference remains a challenge. In this paper, we propose Diff-POI: a Diffusion-based model that samples the user's spatial preference for the next POI recommendation. Inspired by the wide application of diffusion algorithm in sampling from distributions, \method{} encodes the user's visiting sequence and spatial character with two tailor-designed graph encoding modules, followed by a diffusion-based sampling strategy to explore the user's spatial visiting trends. We leverage the diffusion process and its reversed form to sample from the posterior distribution and optimized the corresponding score function. We design a joint training and inference framework to optimize and evaluate the proposed \method{}. Extensive experiments on four real-world POI recommendation datasets demonstrate the superiority of our \method{} over state-of-the-art baseline methods. Further ablation and parameter studies on \method{} reveal the functionality and effectiveness of the proposed diffusion-based sampling strategy for addressing the limitations of existing methods. 
    
\end{abstract}


\setcopyright{acmlicensed}
\acmJournal{TOIS}
\acmYear{2023} \acmVolume{1} \acmNumber{1} \acmArticle{1} \acmMonth{1} \acmPrice{15.00}\acmDOI{10.1145/3624475}

\maketitle

\section{Introduction}

\begin{figure}
    \centering
    \includegraphics[width=\linewidth]{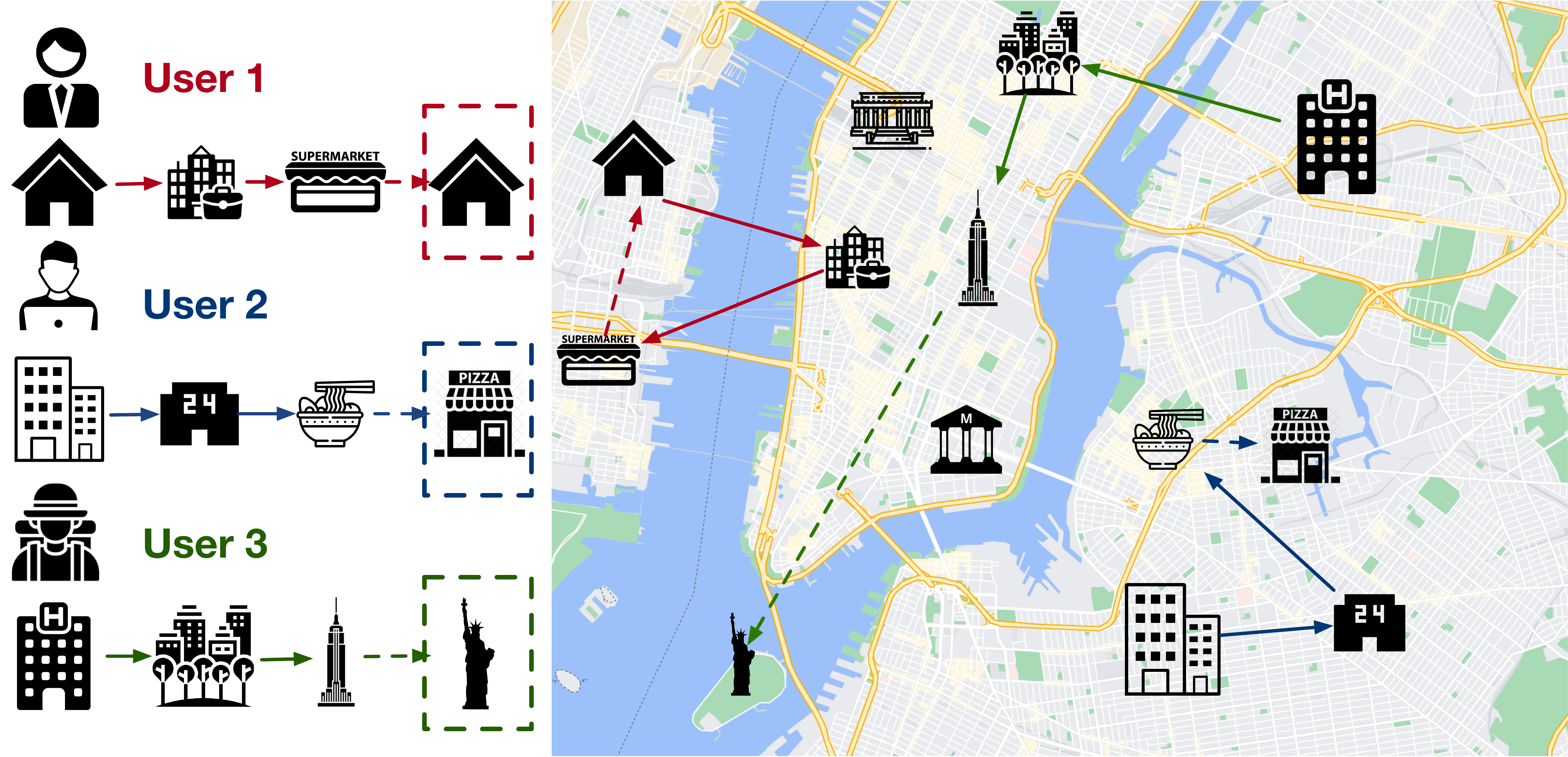}
    \caption{A toy example of different geographical visiting patterns. User 1 usually follows a daily routine and commutes between workplace and home, and would occasionally drop by some spots like supermarkets along the way. User 2 works from home and is interested in exploring restaurants around the dwelling. User 3 has just arrived at the hotel and is about to travel around famous tourist attractions in the city. While the various implicit locational preference lies in historical visits, previous aggregation-based representation would be insufficient for exploring POIs in unfamiliar places.}
    \label{fig:toy}
\end{figure}

The prosperity of Location-Based Social Networks (LBSN) and location-based services apps has sparked significant interest in the development of Point-of-Interest (POI) recommender systems, which provide users with personalized and location-aware recommendations on various services and products. POI recommendation models have become  have become integral to numerous check-in apps, such as Yelp and Foursquare. These location-based apps store user check-ins and provide recommendations for the most suitable POIs, aiming to alleviate the information overload of online platforms. Unlike other sequential recommendation tasks, POI recommendation focuses on capturing the geographical characteristics of the POIs, and how their locations influence user behavior \cite{lian2014geomf,liu2016predicting,yuan2014graph}. 

The success of POI recommendation depends on accurately determining which POI a specific user is most likely to visit based on their previous behavior. This requires joint modeling of both locality features and temporal patterns of the user's visiting sequences. There has been a variety of attempts to find the ideal method to incorporate this spatio-temporal feature, such as distance between POIs \cite{lian2014geomf,feng2015personalized,zhao2016stellar}, or time gap between successive visits \cite{zhao2020go,kong2018hst}. Above that, existing works proposed different methods to encode the observed sequences, such as Markov Chains \cite{cheng2013you,he2016fusing}. Recurrent Neural Networks (RNNs) and their variants are also widely adopted by previous works \cite{zhao2020go,sun2020go,feng2018deepmove,guo2020attentional} and benefited from their capability of capturing the sequential dependencies and long-term preference from visiting sequences. With the development of Graph Neural Networks (GNNs) in recent years, the expressiveness and rich syntax of graph-structured data raises concerns among researchers. Lots of graph-based methods have been proposed \cite{li2021discovering,lim2022hierarchical,wang2022graph,han2020stgcn} to depict the transition and location relationship between POIs. 

Despite the variety of model structures, how to leverage the obtained sequence encodings remains a fundamental problem. The sequential representations contain the temporal and spatial information from the user's previous behavior, yet it is still unclear where the user would move on to the next 
. Some contrastive learning methods are proposed \cite{ju2022kernel} to encourage consistency between the two representations, while other works propose to disentangle the sequential and spatial influences \cite{wang2022learning,qin2023disenpoi}. 

Although the effectiveness of existing works, an important yet common situation is neglected, which is when users aim to explore Point-of-Interest (POI) recommendations in unfamiliar areas. For historical visits, a common practice is to aggregate the corresponding locational representations via a pooling method or through the attention mechanism \cite{lian2020geography,wang2022learning,luo2021stan}. These methods successfully summarize the geographical feature of visiting history and tend to recommend nearby POIs for the user \cite{qin2023disenpoi}. However, they often fail to cater to users who seek to explore novel areas. Figure \ref{fig:toy} illustrates this issue with a toy example, showcasing different types of users and their preferences for diverse visiting tour routes.
User 1 represents a typical conservative visitor, who is more likely to revisit the same POIs or explore familiar areas. User 2 is a curious traveler who enjoys stepping out of their comfort zone occasionally, while User 3 embodies a long-distance tourist who is exploring novel areas within the city.
Despite the diverse visiting patterns among users, aggregation-based representation learning methods can only collect and aggregate historical geographical information and make recommendations based on locational similarity between POIs. 
This simplistic approach may yield promising results for users like User 1 who follow regular routines but can lead to sub-optimal solutions for adventurous users like User 2 and tourists seeking to explore the city like User 3.
An effective method to determine whether to visit novel areas remains unexplored. \R{To tackle this problem, we require a method that bridges the gap between the historical visiting distribution and the target preference distribution. Although research on this inter-distribution relationship is still limited, we can rely on the diffusion generative method, which is a popular theory in the field of generative studies and effectively resolves this problem.}

The rapid development of diffusion-based methods since the Denoising Diffusion Probabilistic Model (DDPM) \cite{sohl2015deep,ho2020denoising} has garnered increasing attention. By utilizing Stochastic Differential Equations (SDEs) to model the diffusion process, the reverse diffusion process can be seen as a Langevin sampling process from a given prior distribution \cite{song2020score}. Further studies on score-matching and Langevin dynamics have provided more insights into the application of existing diffusion algorithms. Notably, these methods have shown significant improvements over traditional approaches in various fields, such as image editing \cite{meng2021sdedit}, natural language processing \cite{li2022diffusion}, and graph generation \cite{vignac2022digress,xu2022geodiff}. The combination of diffusion process and score-based models enables sampling from distributions represented by specific score functions, expanding the application of diffusion-based methods beyond generative tasks.

\R{Recent studies on diffusion models \cite{liu2022flow,song2023consistency,zhao2022egsde} has shed light on the nature of the diffusion process.It empowers generative models with the ability to depict the transportation path from source distributions to target distributions. From the perspective of diffusion process, the overall transition between distributions is decomposed into a series of stochastic process, which can be better captured with specific model structures, such as neural networks. In the context of generative methods, this benefit of the diffusion process aids models in accurately representing the complex sample space compared to traditional methods such as VAEs \cite{rezende2014stochastic,kingma2013auto} and GANs \cite{goodfellow2020generative}. Under this assumption, layers that map between distributions are trained in a step-by-step manner, which fully explores the expressiveness of models.}

\R{In the case of POI recommendation, there exists a similar dilemma in traditional methods, where the relationship between a user's historical visits and their future preferences remains unclear. As mentioned earlier, previous location-based representation learning methods have successfully modeled the geographical features of visiting sequences but lack an effective way to indicate a user's future preference. A straightforward approach would be to apply an aggregation layer to directly transform the historical distribution into a future preference. However, these traditional solutions lack explicit modeling of the subtle relationship between these two distributions.
Thereby, inspired by the idea of Langevin sampling from specific distributions, we propose addressing the aforementioned problem in location-based POI recommendation with a tailored geo-preference diffusion process.}

In a nutshell, we propose \textbf{Diff-POI}: a \textbf{Diff}usion-based model that samples the user's spatial preference for the next \textbf{POI} recommendation. Specifically, we design an attention-based graph encoder to integrate spatio-temporal information with traditional sequential graphs to obtain fine-grained user embeddings. Moreover, a graph convolution module is applied for generating geographical representations for POIs based on their locations. Subsequently, we adopt a context-conducted attention layer to obtain location prototypes for corresponding users. Finally, a score function is applied to model the gradient of the posterior distributions, which are parameterized by location prototypes. From the posterior distributions, we sample the target location embeddings via a reversed diffusion process. The location and user embeddings are jointly considered to make POI recommendations. Empirical studies show the superiority of \method{} against state-of-the-art baseline methods.

To summarize, the contributions of our work are listed as follows: 

\begin{itemize}
    \item We propose an attention-based graph encoder that extends traditional sequential GNNs by integrating extra spatio-temporal information from observed visiting sequences. The novel sequence graph encoder can generate fine-grained embeddings for users.
    \item We propose \method{}, a diffusion-based model that samples from the posterior distribution that reflects the user's geographical preference. With the help of the diffusion process, \method{} is able to fully exploit the potential of the obtained location and user embeddings.
    \item Comprehensive experiments on four real-world LSBN datasets demonstrate not only the effectiveness and robustness of the proposed \method{} against the state-of-the-art baselines but also its capability of depicting different locational visiting patterns.
\end{itemize}

\section{Related Work}
\label{sec::related}

In this section, we will introduce the related works from two aspects, namely the next POI Recommendation, graph neural networks, and SDEs with denoising diffusion.

\subsection{Next POI Recommendation}

The next POI recommendation lies at the core of numerous location-based services, which aim to recommend the most possible POI for the user to visit. Previous attempts manage to integrate the spatio-temporal information with existing recommendation methods, such as matrix factorization \cite{mnih2007probabilistic,rendle2010factorizing} and Markov chains \cite{cheng2013you,he2016fusing}. Further studies extend the model by explicitly considering the temporal or locational similarity between POIs \cite{lian2013collaborative,wang2018exploiting,huang2022empowering}.

With the development of sequential recommendation models and deep learning methods, researchers have put more attention on deep learning-based methods to model the sequential evolution in POI recommendation. Recurrent Neural Networks (RNNs) and their variants are widely used to encode visiting sequences. ST-RNN \cite{liu2016predicting} proposes to use different transition matrices to model contexts from different perspectives in RNNs. STGN \cite{zhao2020go} introduces two extra time and distance gates to LSTM structures to better capture the spatio-temporal gap in visiting sequences. LSTPM \cite{sun2020go} proposes to model the temporal and geographical sequences separately, by encoding POI sequences with a standard LSTM and a geo-dilated LSTM correspondingly. Apart from improving RNN structures, the attention mechanism is also widely adopted by recent works to discover multiple and flexible factors behind visits, such as ARNN \cite{guo2020attentional}. There are also attempts to leverage transformer-like self-attention structure for POI recommendation tasks \cite{yang2022getnext,qin2022next}. GeoSAN \cite{lian2020geography} equips the self-attention layer with a geography encoder to encode locations with GPS coordinates. STAN \cite{luo2021stan} further improves the location encoder by leveraging relative spatio-temporal information in user trajectories.

\subsection{Graph Neural Networks}

The popularity of Graph Neural Networks (GNNs) has seen tremendous success, demonstrating exceptional performance across various graph-related tasks such as node classification \cite{kipf2016semi,ju2023comprehensive,yuan2023learning}, graph classification \cite{luo2022dualgraph,ju2023tgnn,luo2023towards}, and graph clustering \cite{bo2020structural,ju2023glcc}. Their immense potential has propelled the development of widespread applications, including collaborative filtering \cite{wang2022disenctr,qin2023learning}, traffic flow prediction \cite{zhao2023dynamic}, and drug discovery \cite{ju2023few}.
A popular trend in recommender system research is the application of GNN structures. Since the success of graph convolution networks \cite{kipf2016semi}, graph-structured data in POI recommendation tasks have been widely studied \cite{rao2022graph,lim2020stp}. GE \cite{xie2016learning} collects information from various kinds of POI and user graphs. STGCN \cite{han2020stgcn} exploits the idea of the Relational Graph Convolution Network (RGCN) to depict the multiple relationships between users and POIs. Later works like GSTN \cite{wang2022graph} lay their focus on the transition and distance POI graphs. DisenPOI \cite{qin2023disenpoi} further proposes to disentangle the geographical and sequential effects with a contrastive regularization to fully exploit the different factors behind visits. \R{More recent studies \cite{li2021discovering,wang2022graph,wang2022learning,ju2022kernel,qin2023disenpoi} focus on the dual graph encoder structures that are capable of jointly modeling the spatio-temporal influence through the expressive graph structures.}

Generally, existing graph-based methods all suffer from the limitations brought by the location embeddings learned from the POI distance graph, and models tend to recommend nearby POIs as previous researches suggested \cite{qin2023disenpoi,ju2022kernel}.

\subsection{SDEs with Denoising Diffusion}

Diffusion models \cite{sohl2015deep,ho2020denoising} have achieved great success in multiple fields. Further researches reveal the relationship between the denoising diffusion process and score-based methods \cite{song2019generative,song2020score}, which bridge the gap between two generative paradigms. Models are using SDEs to calculate and simulate the Langevin sampling process from specific diffusion process for image generation \cite{song2020score} and are gaining great advantage against traditional generative methods such as GANs \cite{dhariwal2021diffusion}. The diffusion process is also viewed as transportation between distributions and can be used in domains including image edit \cite{zhao2022egsde,meng2021sdedit}, natural language processing \cite{li2022diffusion}, audio generation \cite{kong2020diffwave} and graph generation \cite{vignac2022digress,xu2022geodiff}. However, other applications of the denoising diffusion process remains unexplored.

In a nutshell, the strong capability of denoising diffusion process and the view of Langevin sampling widely extend the application of diffusion-based models. Inspired by its success in other fields, we believe learning SDEs with the denoising diffusion process is the key to exploiting the rich syntax from historical location embeddings.

\section{Preliminary}
\label{sec::definition}

In this section, we first define the notations and formulate the POI recommendation problem, then introduce the main idea of constructing spatio-temporal dual graphs, followed by the basis of the SDE-based diffusion process.

\subsection{Sequential POI Recommendation}

Considering the typical settings of POI recommendation. For the POI set $\mathcal{L}=\{l1,l2,...,l_{|\mathcal{L}|}\}$ and user set $\mathcal{U}=\{u1,u2,...,u_{|\mathcal{U}|}\}$, the task can be formulated as: given a user $u$, its trajectory of POIs and visiting timestamps $H(u)=\{(l_1^u,t_1^u),(l_2^u,t_2^u),...(l_{n}^u,t_{n}^u)\}$, the target is to recommend target POI $l_{n+1}^u$ at current timestamp $t_{n+1}^u$.

In the case of sequential POI recommendation, the temporal and distance gap between visits is an important factor to make recommendations. For POI set $\mathcal{L}$, we can obtain the distance matrix $A_d\in\mathbb{R}^{N\times N}$, where $A_d(i,j)=haversine(l_i,l_j)$ denotes the distance between $l_i$ and $l_j$ in km, calculated by haversine formula.

\subsection{Spatio-temporal Dual Graph}

Now we construct dual graph pair for each visiting trajectory to depict the spatial and temporal relationship between POIs respectively. For a trajectory $H(u)$, we obtain the corresponding transition graph $\mathcal{G}_{u}=\{\mathcal{V}_u,\mathcal{E}_u\}$, where vertex set $\mathcal{V}_u$ includes all the POIs in $H(u)$, each edge $e=\left<l^u_i,l^u_{i+1}\right>\in\mathcal{E}_u$ indicates a successive visit from $l_i^u$ to $l_{i+1}^u$ in user's visiting history $H(u)$. 

To model the locational influence, we obtain the distance graph $\mathcal{G}_d=\{\mathcal{V}_d,\mathcal{E}_d,\mathcal{A}_d\}$ for all POIs, where node set $\mathcal{V}_d=\mathcal{L}$. The distance edge $(l_i,l_j)\in\mathcal{E}_d$ indicates the distance between $l_i$ and $l_j$ is within a specific threshold, e.g. 1km as previous works suggested \cite{wang2022graph} and the edge weight $\mathcal{A}_u$ represents the geographical distance.

\subsection{SDE-based Diffusion Process}

Consider a data point sampled from a specific prior $x_0\sim p_{data}(x)$. Given a series of noise scales $\beta_0,\beta_1,...,\beta_T$, the discrete diffusion process is considered as a Markov Chain with the transition probability and the conditional probability is formulated as: 
\begin{equation}
    p(x_{t}|x_{t-1})=\mathcal{N}(x_t;\sqrt{1-\beta_t}x_{t-1},\beta_tI).
\end{equation}
More generally, we can formulate any discrete one-step diffusion into continuous form and reform the process as the solution to specific It\^{o} SDE \cite{song2020score}:
\begin{equation}
\label{eq:sde_exp_forward}
    \mathrm{d}x=f(g,t)\mathrm{d}t+g(t)\mathrm{d}w,
\end{equation}
through which can we sample arbitrary $x(t)\sim p_t$. As previous research stated \cite{anderson1982reverse}, the reverse process of Eq. \ref{eq:sde_exp_forward} is also a diffusion process that can be formulated as:
\begin{equation}
\label{eq:sde_exp_backward}
    \mathrm{d}x=[f(x,t)-g^2(t)\nabla_x\log p_t(x)]\mathrm{d}t+g(t)\mathrm{d}\overline{w}.
\end{equation}
Where the $\mathrm{d}w$ and $\mathrm{d}\overline{w}$ in Eq. \ref{eq:sde_exp_forward} and Eq. \ref{eq:sde_exp_backward} are sampled independently from standard Wiener process at each step of the SDEs.

With a specified score function $s_\theta(x)$ that is parameterized by a neural network $\theta$, we can estimate the gradient of the log marginal distribution $\nabla_x\log p_t(x)$ and sample from any $p_t$ for the target data point $x_0$ via the reversed SDE in Eq. \ref{eq:sde_exp_backward}.

\section{Methodology}

\label{sec::model}

\begin{figure*}[t!]
    \centering
    \includegraphics[width=\textwidth]{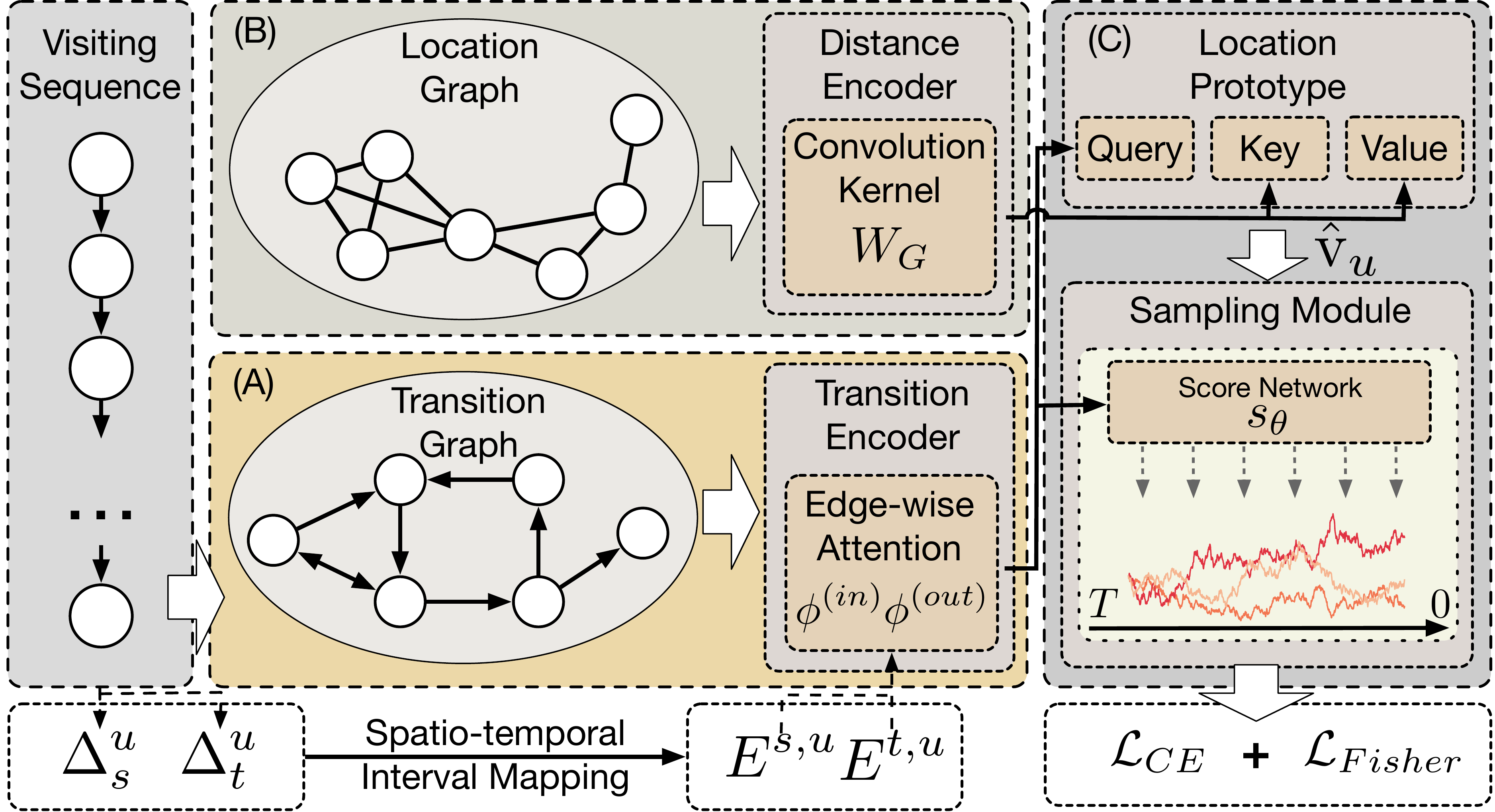}
    \caption{Illustration of the proposed framework \method{}. \method{} is intended for making personalized POI recommendations based on the user's spatial preference and history visiting patterns and is composed of three main parts: (A) a spatio-temporal graph encoder, (B) a distance-based graph convolution encoder, and (C) a diffusion-based sampling module. The details of part (C) will be illustrated in the following sections.}
    \label{fig:framework}
\end{figure*}

In this section, we provide the detailed structure of the proposed \method{}. As illustrated in Figure \ref{fig:framework}, \method{} is composed of three key modules, which are: (A) a spatio-temporal sequence graph encoder that incorporates the distance and time gap between successive visits with sequential graph encodings, (B) a distance-based POI graph encoder that performs graph convolution on the distance graph that yields location embeddings for each POI, and (C) a context-driven diffusion module which obtains user-specific locational prototypes, followed by a diffusion-based sampling strategy to sample the spatial preference for users.

\subsection{User Preference Encoding}

\subsubsection{Spatio-temporal Embedding Layer}

The transition graph $\mathcal{G}_u$ reflects the sequential relationship between POIs within a user's visiting history, making it important for the sequence graph encoder to encode the extra spatio-temporal information between successive visits. An intuitive observation is that the interval between successive visits implies the relationship between these two visits. \R{For instance, a series of closely spaced visits reflects the trajectory of a single trip, suggesting that the user is more likely to continue visiting nearby POIs. Conversely, a long distance between two visits indicates a potential change in the user's locational preference. This distinction calls for an explicit modeling of the intervals between visits from both spatial and temporal perspectives. To achieve this, we propose modeling the temporal and locational intervals using a trainable embedding matrix.} 
For a transition graph $\mathcal{G}_u$ and its corresponding visiting sequence $H(u),|H(u)|=n$, we first obtain the relative spatial and temporal interval matrices $\Delta^u_s,\Delta^u_t\in\mathbb{R}^{n\times n}$:

\begin{equation}
    \Delta^u_s=\begin{bmatrix}
    s^u_{11}\ s^u_{12}\ \dots\ s^u_{1n} \\
    s^u_{21}\ s^u_{22}\ \dots\ s^u_{2n} \\
    \dots\ \dots\ \dots\ \dots \\
    s^u_{n1}\ s^u_{n2}\ \dots\ s^u_{nn}
    \end{bmatrix},\ 
    \Delta^u_t=\begin{bmatrix}
    t^u_{11}\ t^u_{12}\ \dots\ t^u_{1n} \\
    t^u_{21}\ t^u_{22}\ \dots\ t^u_{2n} \\
    \dots\ \dots\ \dots\ \dots \\
    t^u_{n1}\ t^u_{n2}\ \dots\ t^u_{nn}
    \end{bmatrix},
\end{equation}
\R{where $s_{ij}^u=\lfloor \frac{A_d(l_i^u,l_j^u)}{s_{min}^u} \rfloor$ is the relative distance between the $i$-th and $j$-th POI in $H(u)$, normalized by the minimal transition distance in the visiting sequence $s_{min}^u$. We round down the normalized distance to discrete integers so that intervals at different levels can be represented via a list of embedding vectors.} Similarly, the temporal intervals are defined with $t_{ij}^u=\lfloor \frac{t_j^u-t_i^u}{t_{min}^u} \rfloor$.

We then maintain two trainable embedding matrices to represent the intervals respectively, after clipping the relative interval matrices $\Delta_s^u,\Delta_t^u$ with two threshold values $\delta_s,\delta_t$:
\begin{equation}
    s_{ij}^u=\max(s_{ij}^u,\delta_s),\ t_{ij}^u=\max(t_{ij}^u,\delta_t),\ \forall i,j,
\end{equation}
where $\delta_s$ and $\delta_t$ are hyper-parameters. The intervals are further transferred into trainable embedding vectors via interval embedding matrices $E^s\in\mathbb{R}^{\delta_s\times d},E^t\in\mathbb{R}^{\delta_t\times d}$. We can obtain the interval embeddings for arbitrary interval matrices via an interval mapping operation:
\begin{equation}
    e^{s,u}_{ij}=E^s_{s_{ij}^u,:},\ e^{t,u}_{ij}=E^t_{t_{ij}^u,:},
\end{equation}
where $e_{ij}^{s,u},e_{ij}^{t,u} \in\mathbb{R}^d$ are interval embeddings, $d$ represents the embedding size.
 
\subsubsection{Spatio-temporal Sequence Graph Encoder}

\begin{figure*}[t!]
    \centering
    \includegraphics[width=\textwidth]{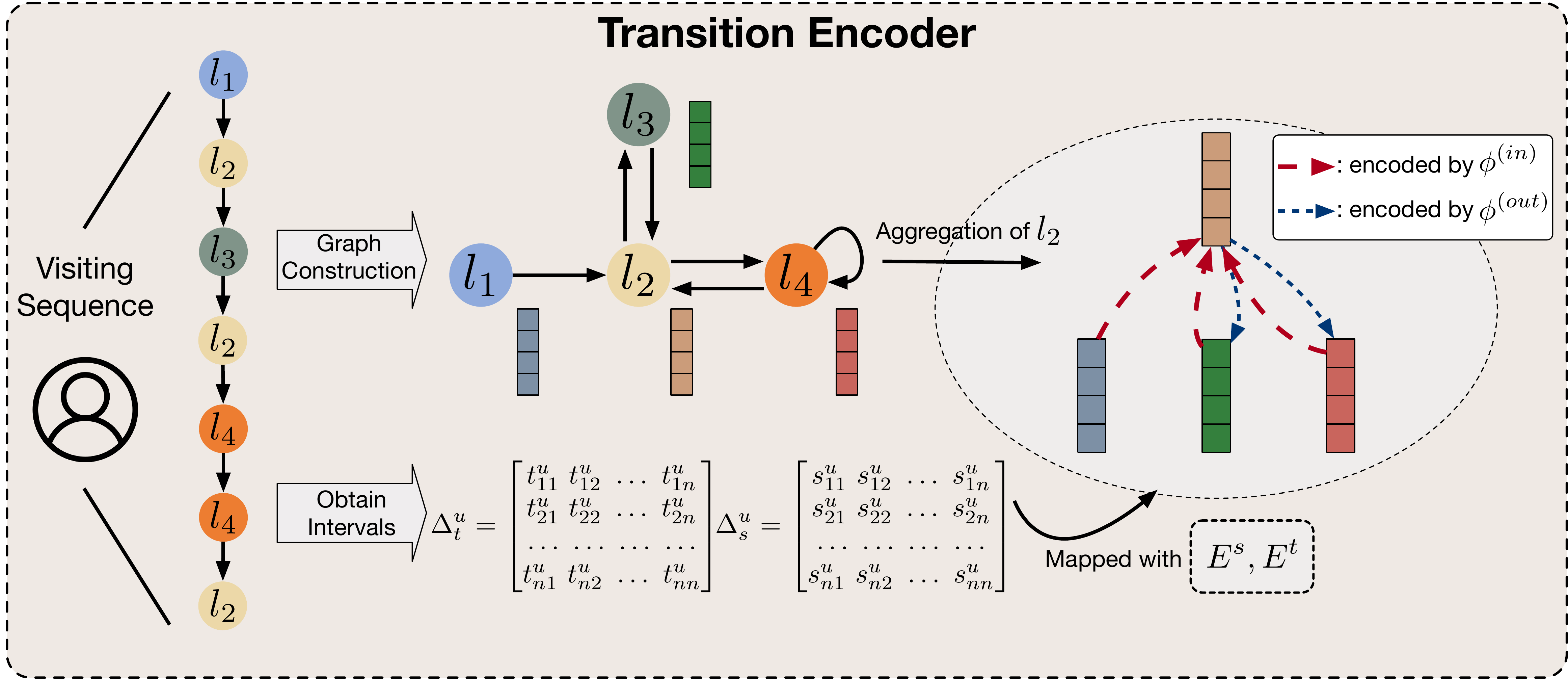}
    \caption{Illustration of the spatio-temporal sequence graph encoder. When generating messages of a POI, we use two projection matrices to calculate the attention coefficients and reaggregate the neighborhood }
    \label{fig:sequence_encoder}
\end{figure*}
Now we design the attention-based graph encoder to process the sequence graphs. For input sequence graph and POI embeddings $e^l_i\in\mathbb{R}^{d}, i=1,2,...,|\mathcal{L}|$, the graph encoder should output the user encoding $x_u\in\mathbb{R}^d$ for the corresponding user $u$. A simple solution is to apply variants of Gate-controlled GNNs (GGNNs) to aggregate the sequence information. However, such a method only receives messages from precursor nodes when generating node representations. Therefore, we propose a tailor-designed attention-based graph encoder to encode spatio-temporal information from both directions on the transition graph. The detailed illustration of the graph encoder is presented in Figure \ref{fig:sequence_encoder}.

Since bi-directional information is equally important for sequential recommendation cases \cite{sun2019bert4rec,qiu2020gag}, we expect the message function could reflect both the feature from previous visits and future choices. Specifically, the message function is applied to receive from incoming and outgoing edges to collect messages from both directions:
\begin{equation}
\label{eq:seq_message}
m_i=m^{(in)}_{i}+m^{(out)}_{i}=\sum_{\left<l_j,l_i\right>\in\mathcal{E}_u} \alpha^{(in)}_{ij}e_j^l+\sum_{\left<l_i,l_k\right>\in\mathcal{E}_u} \alpha^{(out)}_{ik}e_k^l,
\end{equation}
the coefficients $\alpha_{ij}^{(in)}$ and $\alpha_{ik}^{(out)}$ are attention weights of the corresponding neighbors. We calculate them with an edge-wise attention layer, formulated as:
\begin{align}
\begin{cases}
    a^{(in)}_{ij} = \phi^{(in)}(e_i^l\odot e_j^l+e_{ij}^{s,u}+e_{ij}^{y,u}) \\
    a^{(out)}_{ik} = \phi^{(out)}(e_i^l\odot e_k^l+e_{ik}^{s,u}+e_{ik}^{y,u}) \\
    \alpha_{ij} = \mathrm{Softmax}(\mathrm{CONCAT}([a^{(in)}_{ij};a^{(out)}_{ik}])),
\end{cases}
\end{align}
where $\phi^{(in)},\phi^{(in)}:\mathbb{R}^d\rightarrow\mathbb{R}$ are two trainable projection matrices to obtain the attention logits, $\odot$ denotes element-wise product to model the affinity between neighboring nodes.

So far, with the message function defined in Eq. \ref{eq:seq_message}, we can iteratively update the node representation in $\mathcal{G}_u$ with the weighted neighborhood influence, and the outputs noted as $E^h_u=[e_1^h,e_2^h,...,e_n^h]$. To effectively model the user's historical preference, we propose to leverage a self-attention as the readout function for the encoded node representations of sequence graph $\mathcal{G}_u$, noted as $x_u\in\mathbb{R}^d$:
\begin{align}
\label{eq:user_emb}
    x_u=\mathrm{MEAN}\{\mathrm{Attention}(W_Q^hE^h_u,\ W_K^hE^h_u,\ W_V^hE^h_u)\}, \mathrm{with} \\
    \mathrm{Attention}(Q,K,V)\triangleq \mathrm{FFN}(V + \mathrm{Softmax}(\frac{QK^T}{\sqrt{d}})\cdot V)
\end{align}
where $W_Q^h,W_K^h,W_V^h:\mathbb{R}^d\rightarrow\mathbb{R}^d$ are learnable projection matrices, $\mathrm{FFN}$ indicates a layer of feed-forward neural network. While the sequence graph encoder depicts the local interval features between adjacent nodes in $\mathcal{G}_u$, the self-attention readout layer captures the long-term global dependencies between visits to further generate fine-grained sequence embeddings.

\subsection{User-specific Spatial Posterior}

Recall that a POI geographical graph $\mathcal{G}_g$ is constructed based on the distance between POIs, we consider a graph convolution module that encodes $\mathcal{G}_g$ and outputs the location embeddings of POIs, marked as $e^g_i\in\mathbb{R}^d$ for any POI $l_i$. We further obtain the user-specific spatial posterior based on $e^g$ and $x_u$ to sample the locational preference from.

\subsubsection{GNN-based Location Graph Eecoder}
To extract the locational relationship between POIs from $\mathcal{G}_g$, we adopt a graph convolution network \cite{kipf2016semi} as the geographical graph encoder. Specifically, the update function of each layer is built with:
\begin{equation}
    h_{i}^{(l)}=\sum_{j\in\mathcal{N}_i}\frac{w_{ij}}{\sqrt{|\mathcal{N}_i||\mathcal{N}_j|}}W_G^{(l-1)}h_j^{(l-1)},
\end{equation}
where the node representation $h^{0}_i$ is initialized with $e_i^l$ for each node, $W_G^{(l)}$ represents the convolution kernel of $l$-th layer. To dynamically simulate the attenuation of geographical influence as the distance increases, we adapt $w_{ij}=e^{-A_d(l_i,l_j)}$ as the distance edge weight. Benefiting from GCN's capability of aggregating and smoothing neighborhood information, we expect the updated node representation $h^{(l)}$s to reflect the implicit communities and local structures of POIs based on their locational distribution. With the fixed graph topology of $\mathcal{G}_g$ and the number of convolution layers $L$, we readout the locational embeddings $E^g=[e_1^g,e_2^g,...,e_{|\mathcal{L}|}^g]$ with a mean pooling function on the outputs of each layer to capture the high-order connectivity between nodes:
\begin{equation}
\label{eq:geo_embedding}
    e_i^g=\mathrm{MEAN}\{h_i^{(0)},h_i^{(1)},...,h_i^{(l)}\}.\ i=1,2,...,|\mathcal{L}|.
\end{equation}

\subsubsection{User-specific Location Prototype}
Since we have obtained the locational embeddings for POIs, we aim to figure out a method to calculate the probability $p(l=l_i|H(u),\mathcal{G}_g)$ for arbitrary user $u$ to visit $l_i$ according to $u$'s locational preference. To achieve this, we parameterize the probability with a user-specific prototype, formulated as:
\begin{equation}
\label{eq:probability}
p(l|H(u),\mathcal{G}_g)=p(l|\mathrm{v}_u)p(\mathrm{v}_u|H(u),\mathcal{G}_g),
\end{equation}
where $\mathrm{v}_u\in\mathbb{R}^d$ denotes the geographical prototype for the user $u$. The first term of Eq. \ref{eq:probability} can be further interpreted as:
\begin{equation}
    p(l=l_i|\mathrm{v}_u)=\mathrm{sim}(e_i^g,\mathrm{v}_u),
\end{equation}
$\mathrm{sim}(\cdot, \cdot)$ is a similarity function and here we adopt the inner product. Now we only need to sample $\mathrm{v}_u$ from the posterior distribution $p(\mathrm{v}_u|H(u),\mathcal{G}_g)$ to calculate the probability in Eq. \ref{eq:probability}.

We initialize the prototype $\mathrm{v}_u$ to assist and accelerate the sampling process from the posterior. Directly apply mean pooling on visited POIs could be a simple solution, yet it neglects the user's own characteristics. We expect the generated prior could reflect the user's preference when exploring areas and make it able to bring more personalized information.
To achieve this, we leverage the target attention mechanism to initialize the prototype $\mathrm{v}_u$, before further feeding it into the sampling module. In particular, the user encoding $x_u$ is denoted as the key vector to calculate the relativity of each POIs in $u$'s visiting history. The initial of $\mathrm{v}_u$ is formulated as:
\begin{equation}
    \hat{\mathrm{v}}_u=\mathrm{Attention}(W_Q^gx_u,W_K^gE^g_{H(u)},W_V^gE^g_{H(u)}),
\end{equation}
where $W_Q^g,W_K^g,W_V^g$ are trainable parameters. \R{$\hat{\mathrm{v}}_u$ shows $u$'s historic preference on different locations of the city. }

\subsection{Diffusion-based Sampling Module}

\begin{figure*}[t!]
    \centering
    \includegraphics[width=\textwidth]{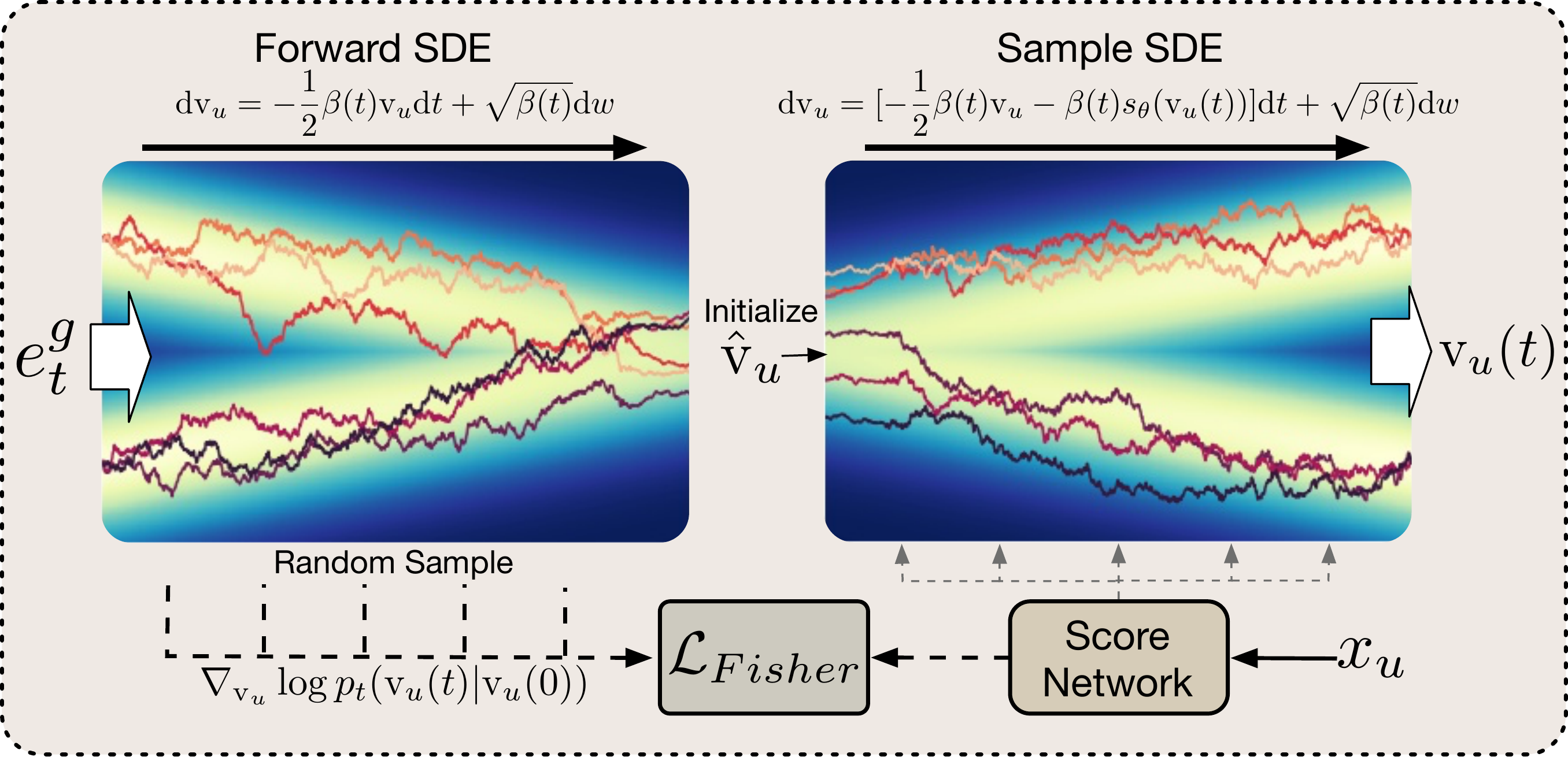}
    \caption{Illustration of the diffusion-based sampling strategy We use the reverse (sample) SDE, which is controlled by the user's historical representation, to sample from the user-specific posterior to generate personalized locational preferences. The marginal probabilities are sampled to optimize the score function.}
    \label{fig:sampling_module}
\end{figure*}

\label{sec::model}

\R{So far we have obtained a historic locational preference $\hat{\mathrm{v}}_u$ from previous sections. However, directly use $\mathrm{v}_u$ to estimate the conditional probability $p(\mathrm{v}_u|H(u),\mathcal{G}_g)$ in Eq. \ref{eq:probability} would lead to sub-optimal results, since it doesn't fully reflect user's future locational preference. We address this issue by the diffusion-based sampling module in the following section.}

To sample from the posterior distribution in Eq. \ref{eq:probability}, a common practice is to use stochastic gradient Langevin dynamics \cite{welling2011bayesian} to iteratively update the sampled parameter. However, the gradient of log probability density, i.e. the score of the posterior, in our case is intractable to calculate. Inspired by the Score-based SDEs \cite{song2020score}, we propose to estimate the score with a score function and model the sampling process with the Variance Preserving (VP) SDE. Though diffusion processes with other perturbation kernels may achieve better performance, experimental results demonstrate that VP-SDE already shows promising results in modeling spatial preferences. Details of the proposed sampling module are illustrated in figure \ref{fig:sampling_module}.

Additionally, the proposed diffusion-based sampling strategy can be adapted to any other POI recommendation frameworks with explicit POI geographical encoders and used for generating an expressive locational prototype with enriched information on the user's spatial performance.

\subsubsection{Forward SDE for Denoising Diffusion}

The aim of the sampling module is to sample the locational embedding of target POI from the user-specific spatial posterior, marked with $e_t^g$ which represents the location embedding of the ground truth target POI $l_t$. In other words, the continuous diffusion process from time $0\sim T$ starts with $\mathrm{v}_u(0)=e_t^g$ and ends with $\mathrm{v}_u(T)=\hat{\mathrm{v}}_u$. Specifically, we model the diffusion process with the following SDE:
\begin{equation}
\label{eq:sde_forward}
    \mathrm{dv}_u=-\frac{1}{2}\beta(t)\mathrm{v}_u\mathrm{d}t+\sqrt{\beta(t)}\mathrm{d}w,
\end{equation}
here we parameterize the noise schedule $\beta(t)=\beta_{min} + \frac{(\beta_{max}-\beta_{min})t}{\beta_{min}T}$ with hyper-parameters $\beta_{min},\beta_{max}$, and $w$ is sampled from a standard Wiener Process at each step of the SDE.

So far, by solving the It\^{o} SDE defined by Eq. \ref{eq:sde_forward} staring from the target data point $\mathrm{v}_u(0)=e_t^g$, we can obtain the perturbed data, which in our case is the denoised spatial posterior.

\subsubsection{Reversed SDE for Preference Sampling}
The sampling process of the location prototype $\mathrm{v}_u$ from the spatial posterior $p(\mathrm{v}_u|H(u)\mathcal{G}_g)$ is equivalent to the reverse process of Eq. \ref{eq:sde_forward}. As previous research concluded \cite{anderson1982reverse}, the reverse SDE is formulated with:
\begin{equation}
\label{eq:sde_backward}
\mathrm{d}\mathrm{v}_u=[-\frac{1}{2}\beta(t)\mathrm{v}_u-\beta(t)\nabla_{\mathrm{v}_u}\log p_t(\mathrm{v}_u)]\mathrm{d}t+\sqrt{\beta(t)}\mathrm{d}\overline{w}.
\end{equation}

Specifically, the spatial preference for $u$ can be sampled with the reversed SDE in Eq. \ref{eq:sde_forward} and we are expecting that the sampled $\mathrm{v}_u(0)$ should be similar to the location embedding of the ground-truth target POI, i.e. $e_t^g$, thus the score function we used should be able to finely estimate real scores.

However, note that the gradient of the log probability, i.e. the score, is intractable to calculate, thus we estimate the score with a score network $s_\theta(\mathrm{v}_u(t))$ parameterized by $\theta$:
\begin{equation}
\label{eq:score_fn}
    s_\theta(\mathrm{v}_u(t))=\mathrm{MLP}(\mathrm{CONCAT}(\mathrm{v}_u(t), x_u)).
\end{equation}
We model the scoring network with a multi-layer perceptron and take the concatenation of user historical embedding and the prototype as input. By introducing the user context condition $x_u$ to the score function, the sampling process is finely controlled by the user's historical behavior, which ensures the sampled prototype fits the user-specific location preference.

The score function is optimized by the Fisher divergence between $s_\theta$ and the score of real samples as previous work suggested \cite{song2019generative}:
\begin{equation}
\label{eq:fisher_loss}
    \mathcal{L}_{Fisher}=\mathbb{E}_{t}[\lVert s_\theta(\mathrm{v}_u(t))-\nabla_{\mathrm{v}_u}\log p_t(\mathrm{v}_u(t)|\mathrm{v}_u(0)) \rVert_2^2],
\end{equation}
where $t\sim U(0,T)$ is uniformly sampled. With $\mathcal{L}_{Fisher}$ as an auxiliary loss, the $s_\theta$ is optimized to better depict the gradient field of the posterior distribution for any given user.

\R{To summarize, we have developed a training and inference framework for sampling personalized geographical preferences for individual users, taking inspiration from the diffusion sampling process. By leveraging the diffusion process, our tailored sampling module has an advantage over other existing generative methods such as VAEs and GANs. It allows our model to learn a detailed transition path from the user's historical preferences to the target distribution. Therefore the sampled preference $\mathrm{v}_u(0)$ can comprehensively represent user's future locational preference.}

\subsection{Model Inference and Optimization}
\subsubsection{Target POI Prediction}

Recall that we have obtained two representative embeddings, i.e. $x_u$ and $\mathrm{v}_u(0)$, that reflect a specific user's transition history and spatial preference respectively. To comprehensively consider the user's preference for the next-to-visit POI, the probability for user $u$ to visit POI $l_i$ is formulated with:
\begin{equation}
    \hat{y}_{i}^u=\mathrm{Softmax}(\alpha e_i^lx_u^T + (1-\alpha) e_i^g(\mathrm{v}_u(0))^T),
\end{equation}
where $\alpha$ is a weight coefficient that reflects the importance of the two similarity terms. By default, we set $\alpha=0.5$ and value these two factors with equal importance.

\subsubsection{Model Optimization}

Given the user's visiting history $H(u)$ and the corresponding ground truth target POI $y_{i}^u$, the model is optimized with a supervised cross-entropy loss:
\begin{equation}
\label{eq:total_loss}
    \mathcal{L}_{CE}=-\frac{1}{|\mathcal{U}|}\sum_{u\in\mathcal{U}}y_i^u\log(\hat{y}_i^u)+\lambda\lVert \Theta \rVert_2^2,
\end{equation}
where $\Theta$ represents model parameters and $\lambda$ is the corresponding weight for the L2 regularization term.

Recall that the score function in the diffusion-based sampling module is optimized via an extra Fisher loss, the overall loss function of \method{} is further written as:
\begin{equation}
    \mathcal{L}_{Total}=\mathcal{L}_{CE}+\gamma\mathcal{L}_{Fisher},
\end{equation}
where $\gamma$ is a hyper-parameter to balance the optimization targets.

We summarize the overall training and inference process in Algorithm \ref{alg:1}

\begin{algorithm2e}[t]
\LinesNumberedHidden
\newcounter{algoline}
\newcommand\Numberline{\refstepcounter{algoline}\nlset{\thealgoline}}
\SetNlSkip{0em}
\SetAlgoNlRelativeSize{0}
\caption{Learning Algorithm of \method{}}
\label{alg:1}
\KwIn{POI set $\mathcal{L}$, User set $\mathcal{U}$, Sampling step num}
\SetKwInput{kwInit}{Initialize}
\kwInit{Distance matrix $A_g$; Model parameters}
Construct $\mathcal{G}_{g}$ based on $A_g$\;
\While{not converged}{
Sample a visiting sequence $H(u)=\{(l_1^u,t_1^u),(l_2^u,t_2^u),...(l_{n}^u,t_{n}^u)\}$ and target POI $l_t$\;
Construct sequential graph $\mathcal{G}_u$ and interval matrices $\Delta_s^u,\Delta_t^u$\;
Encode $\mathcal{G}_u$ to the user encoding $x_u$\;
Obtain initial location prototype $\hat{\mathrm{v}}_u$\;
Initialize $\mathrm{v}_u(0)\leftarrow\hat{\mathrm{v}}_u$\;
\For{$i$ in range stepnum}{
    Sample $\mathrm{d}w$ from standard Wiener process\;
    Calculate $\mathrm{dv}_u$ via Eq. \ref{eq:sde_backward}\;
    Update $\mathrm{v}_u(i)\leftarrow\mathrm{v}_u(i-1)+\mathrm{dv}_u$\;
}

Inference preference score $\hat{y}_i^u$\;
\If{Train with Fisher loss}{
    Sample $t\sim U(0,T)$\;
    Calculate Fisher loss $\mathcal{L}_{Fisher}$\;
}
Calculate and optimize model with overall loss $\mathcal{L}_{Total}$\;
}
\end{algorithm2e}

\R{
\smallskip\textbf{Complexity Analysis.}
In the previous sections, we defined the overall propagation algorithm, which consists of two main parts: (i) the graph encoders responsible for generating sequence embeddings and geographical prototypes, and (ii) the diffusion sampling module used to sample the user's spatial preference.

For (i), assume having a visiting sequence with $n$ POIs, the complexity of the transition graph encoder is primarily determined by the number of edges and the size of node representations. Computing the locational POI embedding involves edge propagation on the geographical graph $\mathcal{G}_g$, and its complexity is related to the edge size $|\mathcal{E}_g|$. The attention module complexity arises from the softmax normalization applied along the sequence. To summarize, the time complexity can be calculated as follows:
\begin{equation}
    O(nd + |\mathcal{E}_g|d + n^2d)=O((|\mathcal{E}_g|+n^2)d).
\end{equation}

For (ii), forward and reversed diffusion process requires the same computational complexity, which is defined by the number of sampling steps and the complexity of the scoring network, formulated as $O(Td)$.

In summary, the overall computational consumption of  \method{} can be estimated by considering the complexities of both parts (i) and (ii):
\begin{equation}
    O((|\mathcal{E}_g|+n^2+T)d).
\end{equation}
}
\section{Experiment}

To evaluate the performance of our \method{} and the effectiveness of the proposed ideas, we conduct a series of experiments and empirical studies on four real-world LBSN datasets. The aim of the experiments is to answer the following research questions.

\textbf{RQ1: }
Compared with the current state-of-the-art baseline methods, how well does the proposed \method{} perform? Is sampling the user's spatial preference with the diffusion process an effective method for POI recommendation? 

\textbf{RQ2: }
How do the proposed sequence graph encoder and the diffusion-based sampling module boost the performance of \method{}? Is the model sensitive to hyper-parameters?

\textbf{RQ3: }
Can \method{} correctly clarify different types of spatial preference? How does the diffusion-based sampling strategy respond to different users?

\subsection{Datasets and Experimental Setup}

\subsubsection{Evaluation Datasets}
\begin{table}
\centering
\caption{Descriptive statistics of the used datasets.}
\label{tab:statics}
\setlength{\tabcolsep}{3pt}
\begin{tabular}{ccccc} 
\toprule 
\textbf{Dataset} & \#User & \#POI & \#Interaction & \#Avg. Visit \\
\midrule
Gowalla & 10162 & 24237 & 456820 & 44.95 \\
SIN & 2321 & 5596 & 194108 & 83.63 \\
TKY & 2293 & 15177 & 494807 & 215.79 \\
NYC & 1083 & 9989 & 179468 & 165.71 \\
\bottomrule 
\end{tabular}
\end{table}
We evaluate the proposed \method{} on four datasets collected from two real-world check-in platforms, namely \textbf{Gowalla}\footnote{{\url{http://snap.stanford.edu/data/loc-gowalla.html}}} and \textbf{Foursquare}\footnote{\url{https://sites.google.com/site/yangdingqi/home/foursquare-dataset}}.
The Gowalla dataset contains users' check-in records on the Gowalla website from February 2009 to October 2010. The Foursquare dataset includes three subsets, which are collected from Singapore, Tokyo, and New York respectively. The Singapore subset is collected from August 2010 to July 2011, and the rest two subsets are from 12 April 2012 to 16 February 2013.

Following previous work \cite{wang2022learning}, we process the aforementioned datasets with the 5-core cleaning strategy, which means the users and POIs with less than 5 visits are filtered out of the dataset. The detailed statistics of the used datasets are listed in Table \ref{tab:statics}. All the check-in records are sorted chronologically and split by the ratio of 80\%,10\%,10\% into train, valid and test set respectively, which is a common practice in next POI recommendation \cite{wang2022learning,sun2020go}. We train and tune all models on the train set, and choose the best epoch on the validation set as the model to be tested on the test set and report the experimental results. 

\subsubsection{Compared Baselines}
To comprehensively evaluate and demonstrate the effectiveness of \method{}, we compare its performance with the following baseline methods from three aspects: (a) traditional interaction-based user-item recommendation methods, (b) general graph-based recommendation methods (c) sequence-based POI recommendation methods that considers spatial factors in recommendation, and (d) graph-based POI recommendation methods. The compared baselines and their brief introducions are listed as follows:
\begin{itemize} 
    \item \textbf{(a)} MF \cite{mnih2007probabilistic}: It is a classical collaborative filtering method that based on low-rank matrix factorization and optimized with gradient decent method.
    \item \R{\textbf{(a)} GRU4Rec \cite{hidasi2015session}: It is one of the most classical methods that applies Recurrent Neural Networks (RNNs) into sequential recommendation tasks.}
    \item \textbf{(b)} LightGCN \cite{he2020lightgcn}: It is a CF method that adopts graph convolution networks to node representation learning for recommendation.
    \item \textbf{(b)} DGCF \cite{wang2020disentangled}: It is a variant of LightGCN that proposes to disentangle the node representations for better performance.
    \item \R{\textbf{(b)} SR-GNN \cite{wu2019session}: It is a transition graph-based method that leverages Gated Graph Neural Networks to predict next item.}
    \item \textbf{(c)} GeoIE \cite{wang2018exploiting}: It is a classical POI recommendation method that learns geographical and interaction influence separately.
    \item \textbf{(c)} LSTPM \cite{sun2020go}: It is a state-of-the-art LSTM-based POI recommendation method that models the visiting trajectories with a nonlocal network and a geo-dilated LSTM.
    \item \textbf{(c)} STAN \cite{luo2021stan}: It is a state-of-the-art attention-based model that collect sequetial information with a spatio-temporal attention network with consideration of personalized item frequency.
    \item \textbf{(d)} SGRec \cite{li2021discovering}: It is a GNN-based POI recommendation method that propose a Seq2graph augmentation to propogate collaborative signals and learn effective sequential patterns.
    \item \textbf{(d)} DRAN \cite{wang2022learning}: It is the state-of-the-art GNN-based method that leverages a disentangled representation-enhanced attention network for next POI recommendation.
\end{itemize}

\subsubsection{Evaluation Protocal}
We adopt Recall and Normalized Discounted Cumulative Gain (a.k.a NDCG) with a cutoff at top-K-rated items as the evaluation metric, which is a common practice. Specifically, K is enumerated from $\{2, 5, 10\}$ for Recall and NDCG. Though some previous works choose to evaluate the model with a certain amount of negative samples, we choose all the POIs as the candidate POI at the evaluation stage to get stable and more convicing results.

\begin{table}
\centering
\caption{The test results of \method{} and all baselines, where R@K and N@K are short for Recall@K and NDCG@K. The highest performance is emphasized with bold font and the second highest is marked with underlines. \textbf{$\star$} indicates that \method{} outperforms the best baseline model at the significance level with a p-value<0.05 level of unpaired t-test.}
\label{tab:overall}
\renewcommand{\arraystretch}{1.1}
\resizebox{\linewidth}{!}{
\begin{tabular}{c|cc|ccc|ccc|cc|c}
\toprule 
\multicolumn{12}{c}{\textbf{Gowalla}} \\
\bottomrule \hline \hline
Metrics & MF & \R{GRU4Rec} & LightGCN & DGCF & \R{SR-GNN} &  GeoIE & LSTPM & STAN & SGRec & DRAN & \method{}\\

\hline
Recall@2 & 0.0414 & 0.1328 & 0.1102 & 0.1147 & 0.1531 & 0.1767 & 0.1904 & \underline{0.2195} & 0.1932 & 0.2133 & \textbf{0.2635\textsuperscript{$\star$}} \\
Recall@5 & 0.0869 & 0.1754 & 0.1348 & 0.1396 & 0.1978 & 0.2123 & 0.2049 & 0.2364 & 0.2286 & \underline{0.2466} & \textbf{0.3086\textsuperscript{$\star$}} \\
Recall@10 & 0.1473 & 0.2164 & 0.1699 & 0.1721 & 0.2384 & 0.2436 & 0.2618 & 0.2994 & 0.2718 & \underline{0.3056} & \textbf{0.3421\textsuperscript{$\star$}} \\
\hline
NDCG@2 & 0.0428 & 0.1124 & 0.0914 & 0.0995 & 0.1402 & 0.1670 & 0.1431 & 0.1917 & 0.1573 & \underline{0.2038} & \textbf{0.2527\textsuperscript{$\star$}} \\
NDCG@5 & 0.0810 & 0.1367 & 0.1162 & 0.1319 & 0.1621 & 0.1829 & 0.1588 & 0.2152 & 0.1662 & \underline{0.2168} & \textbf{0.2716\textsuperscript{$\star$}} \\
NDCG@10 & 0.1078 & 0.1449 & 0.1432 & 0.1561 & 0.1751& 0.1929 & 0.1742 & 0.2268 & 0.1704 & \underline{0.2367} & \textbf{0.2872\textsuperscript{$\star$}} \\
\bottomrule 
\end{tabular}
}

\resizebox{\linewidth}{!}{
\begin{tabular}{c|cc|ccc|ccc|cc|c}
\toprule 
\multicolumn{12}{c}{\textbf{Singapore}} \\
\bottomrule \hline \hline
Metrics & MF & \R{GRU4Rec} & LightGCN & DGCF & \R{SR-GNN}  & GeoIE & LSTPM & STAN & SGRec & DRAN & \method{}\\

\hline
Recall@2 & 0.1103 & 0.1775 & 0.1606 & 0.1798 & 0.2406 & 0.2521 & 0.2704 & 0.2634 & 0.3058 & \underline{0.3086} & \textbf{0.3455\textsuperscript{$\star$}} \\
Recall@5 & 0.1766 & 0.2291 & 0.2144 & 0.2040 & 0.2715 & 0.2867 & 0.3253 & 0.2940 & 0.3507 & \underline{0.3554} & \textbf{0.3906\textsuperscript{$\star$}} \\
Recall@10 & 0.2059 & 0.2774 & 0.2419 & 0.2285 & 0.3080 & 0.3162 & 0.3791 & 0.3280 & 0.3884 & \underline{0.3921} & \textbf{0.4294\textsuperscript{$\star$}} \\
\hline
NDCG@2 & 0.0744 & 0.1652 & 0.1720 & 0.1973 & 0.2272 & 0.2428 & 0.2610 & 0.2831 & 0.2274 & \underline{0.2972} & \textbf{0.3316\textsuperscript{$\star$}}  \\
NDCG@5 & 0.0791 & 0.1884 & 0.1789 & 0.2162 & 0.2815 & 0.2583 & 0.2697 & 0.2995 & 0.2697 & \underline{0.3175} & \textbf{ 0.3518\textsuperscript{$\star$}}  \\
NDCG@10 & 0.0868 & 0.2040 & 0.1892 & 0.2269 & 0.2549 & 0.2679 & 0.2749 & 0.2892 & 0.2916 & \underline{0.3297} & \textbf{0.3643\textsuperscript{$\star$}} \\
\bottomrule 
\end{tabular}
}

\resizebox{\linewidth}{!}{
\begin{tabular}{c|cc|ccc|ccc|cc|c}
\toprule 
\multicolumn{12}{c}{\textbf{Tokyo}} \\
\bottomrule \hline \hline
Metrics & MF & \R{GRU4Rec} & LightGCN & DGCF & \R{SR-GNN}  & GeoIE & LSTPM & STAN & SGRec & DRAN & \method{}\\

\hline
Recall@2 & 0.2896 & 0.4203 & 0.3917 & 0.4204 & 0.4600 & 0.4975 & 0.5029 & 0.5105 & 0.5091 & \underline{0.5225} & \textbf{0.6031\textsuperscript{$\star$}} \\
Recall@5 & 0.3141 & 0.4878 & 0.4473 & 0.4913 & 0.5257 & 0.5408 & 0.5513 & 0.5489 & 0.5488 & \underline{0.5570} & \textbf{0.6401\textsuperscript{$\star$}} \\
Recall@10 & 0.3866 & 0.5205 & 0.4936 & 0.5310 & 0.5458 & 0.5726 & 0.5795 & 0.6167 & 0.6173 & \underline{0.6210} & \textbf{0.6681\textsuperscript{$\star$}}\\
\hline
NDCG@2 & 0.1984 & 0.4109 & 0.4207 & 0.4299 & 0.4537 & 0.4842 & 0.4724 & 0.4990 & 0.4715 & \underline{0.5089} & \textbf{0.5886\textsuperscript{$\star$}} \\
NDCG@5 & 0.2071 & 0.4373 & 0.4354 & 0.4376 & 0.4788 & 0.5037 & 0.4881 & 0.5264 & 0.4905 & \underline{0.5390} & \textbf{0.6062\textsuperscript{$\star$}} \\
NDCG@10 & 0.2327 & 0.4520 & 0.4403 & 0.4558 & 0.4908 & 0.5141 & 0.4962 & 0.5554 & 0.5112 & \underline{0.5602} & \textbf{0.6160\textsuperscript{$\star$}} \\
\bottomrule 
\end{tabular}
}

\resizebox{\linewidth}{!}{
\begin{tabular}{c|cc|ccc|ccc|cc|c}
\toprule 
\multicolumn{12}{c}{\textbf{New York City}} \\
\bottomrule \hline \hline
Metrics & MF & \R{GRU4Rec} & LightGCN & DGCF & \R{SR-GNN}  & GeoIE & LSTPM & STAN & SGRec & DRAN & \method{} \\

\hline
Recall@2 & 0.2361 & 0.4393 & 0.3789 & 0.3368 & 0.5052 & 0.5655 & 0.5754 & \underline{0.6027} & 0.5734 & 0.5859 & \textbf{0.6589\textsuperscript{$\star$}} \\
Recall@5 & 0.2792 & 0.4870 & 0.4363 & 0.4167 & 0.5465 & 0.5994 & 0.6020 & \underline{0.6358} & 0.6175 & 0.6253 & \textbf{0.6755\textsuperscript{$\star$}} \\
Recall@10 & 0.3046 & 0.5238 & 0.4519 & 0.5653 & 0.5951 & 0.6251 & 0.6312 & \underline{0.6533} & 0.6424 & 0.6478 & \textbf{0.6884\textsuperscript{$\star$}} \\
\hline
NDCG@2 & 0.2002 & 0.4252 & 0.3819 & 0.4847 & 0.5191 & 0.5629 & 0.5524 & \underline{0.5887} & 0.5490 & 0.5702 & \textbf{ 0.6536\textsuperscript{$\star$}}   \\
NDCG@5 & 0.2318 & 0.4467 & 0.3867 & 0.4922 & 0.5319 & 0.5716 & 0.5596 & \underline{0.6092} & 0.5559 & 0.5881 & \textbf{0.6616\textsuperscript{$\star$}}   \\
NDCG@10 & 0.2426 & 0.4587 & 0.3903 & 0.5041 & 0.5495 & 0.5805 & 0.5681 & \underline{0.6124} & 0.5613 & 0.5956 & \textbf{0.6745} \\
\bottomrule 
\end{tabular}
}

\end{table}

\subsubsection{Implementation Detail}
We implement the proposed \method{} and all the baseline models in Pytorch. The implementations of baseline methods are based on the released open-sourced projects or from the authors. The embedding size of all models is fixed to 64, the learning rate is fixed as $lr=0.001$ and all models are optimized with Adam Optimizer. For \method{} and DRAN, the distance-based POI graph is constructed with a 1km distance threshold. For \method{}, we set the hyper-parameters with $\alpha=0.5, \gamma=0.2, \lambda=10^{-3}$. The dropout rate is searched from $\{0.1, 0.2, 0.3, 0.4\}$. The two interval thresholds $\delta_s,\delta_t$ are fixed to 256 for \method{}, DRAN, and STAN. For all sequence-based methods, we preserve the latest 100 visits for sequences with more than 100 POIs. For the diffusion-based sampling module in \method{}, we adopt a step size of $0.01$, which means the reversed SDE for the sampling module would be solved from time step 0 to 1 with $\mathrm{d}t=0.01$ at each step. Particularly, we adopt an early-stop mechanism when training and evaluating models with the patience of 10.
The implementation of \method{} is publicly available at \url{https://github.com/Yifang-Qin/Diff-POI}.

\subsection{Overall Comparison (RQ1)}

We conduct the overall experiments for the aforementioned baselines on four datasets and the experimental results are reported in table \ref{tab:overall}. We make the following observations:

\begin{itemize} 
    \item Compared with traditional recommendation methods like MF, models that incorporate locational factors achieve significant improvements. The result demonstrates the importance of explicitly considering the geographical influence in POI recommendation. Moreover, models that leverage the time and distance intervals between successive visits (LSTPM, STAN, DRAN, and \method{}) outperform the regular location-based baselines, which implies that the key to optimal POI recommendation lies in the spatio-temporal features of transition history.
    \item \R{Graph-based methods that leverage the transition graphs (SR-GNN, SGRec, DRAN, and \method{}) have achieved more promising results among all datasets. The observation reflects the effectiveness and rich syntax of graph-structured data in depicting sequential patterns}. With the help of different expressive GNNs, the models are empowered with the capability of modeling the high-order similarity between POIs revealed from sequences.
    \item The proposed \method{} achieves significant improvements on all four datasets compared with the current state-of-the-art baseline methods. In particular, the performance gain on Recall@10 is improved over the best baseline by 11.9\%, 9.5\%, 7.6\%, and 5.3\% on four datasets respectively, and the gain w.r.t. NDCG@10 by 10.8\%, 10.5\%, 9.9\%, and 10.1\%. The result shows the effectiveness of the proposed \method{} to encode user transition graphs with a spatio-temporal graph encoder and a diffusion-based spatial preference sampling module. The attention-based graph encoder depicts the intervals between successive visits and generates fine-grained user preference embeddings specifically, and the diffusion-based sampling strategy samples the corresponding locational embedding to reveal the spatial trends for the next-to-visit POI to recommend.
\end{itemize}

\subsection{Ablation and Parameter Study (RQ2)}

In this section, we explore the functionality and effectiveness of different modules in the proposed \method{} via a series of ablation studies, as well as \method{}'s sensitivity to the hyper-parameters by parameter studies.

\subsubsection{Funcitonality of Sequence Graph Encoder}

\begin{figure}[t]
\centering

\begin{subfigure}{0.4\linewidth}
    \includegraphics[width=\linewidth]{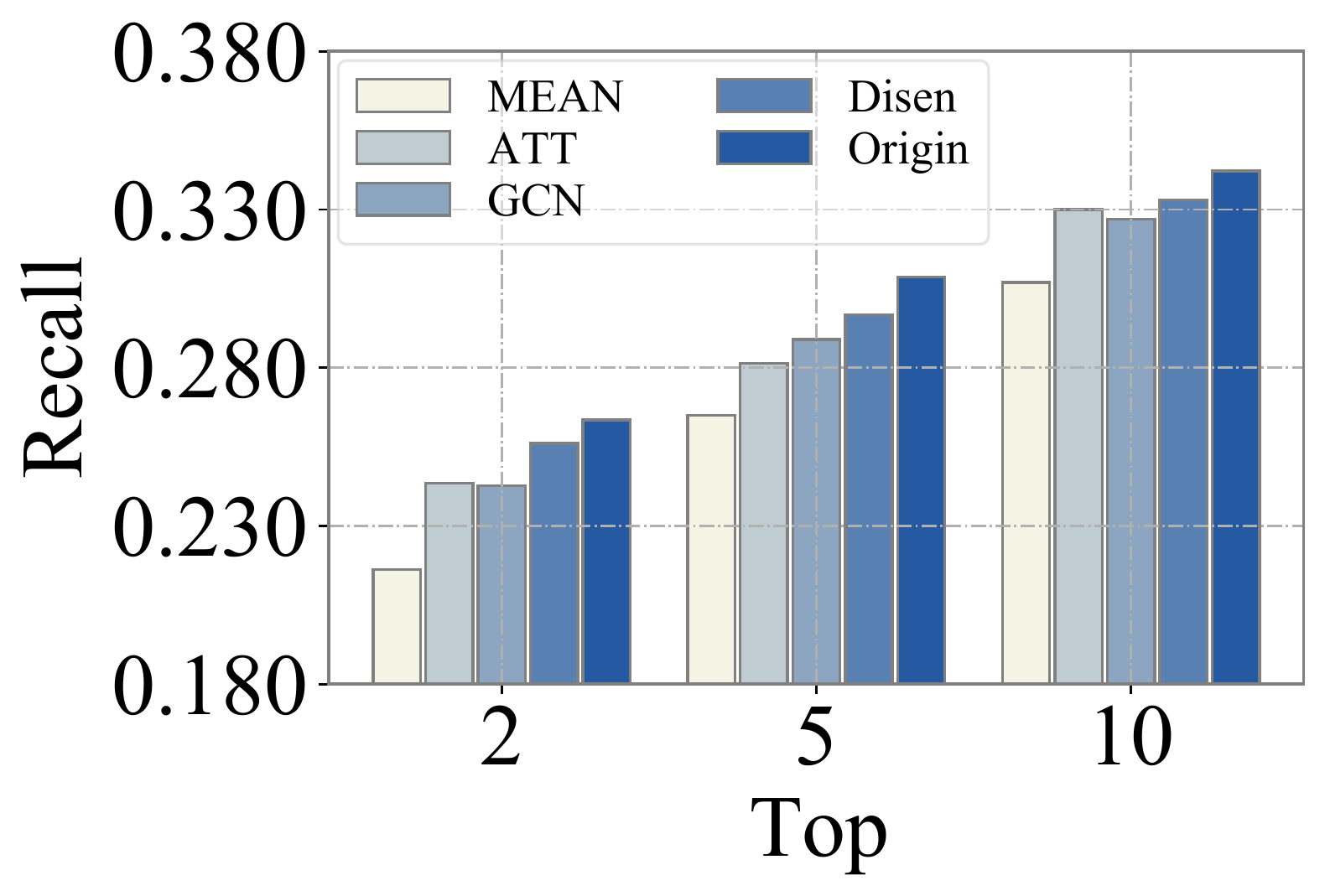}
\caption{Results on \textbf{Gowalla}}
\end{subfigure}
\begin{subfigure}{0.4\linewidth}
    \includegraphics[width=\linewidth]{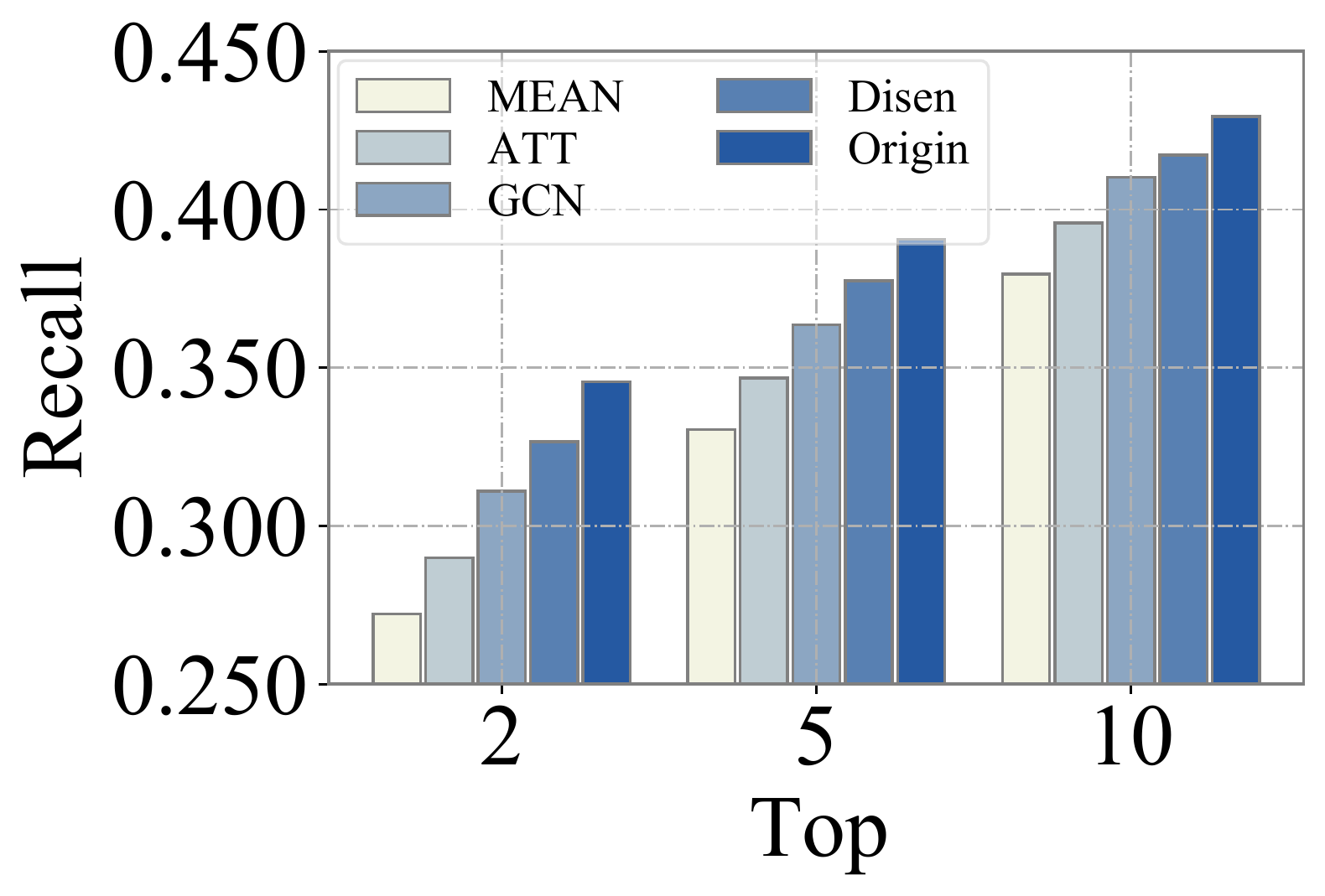}
\caption{Results on \textbf{Singapore}}
\end{subfigure}


\begin{subfigure}{0.4\linewidth}
    \includegraphics[width=\linewidth]{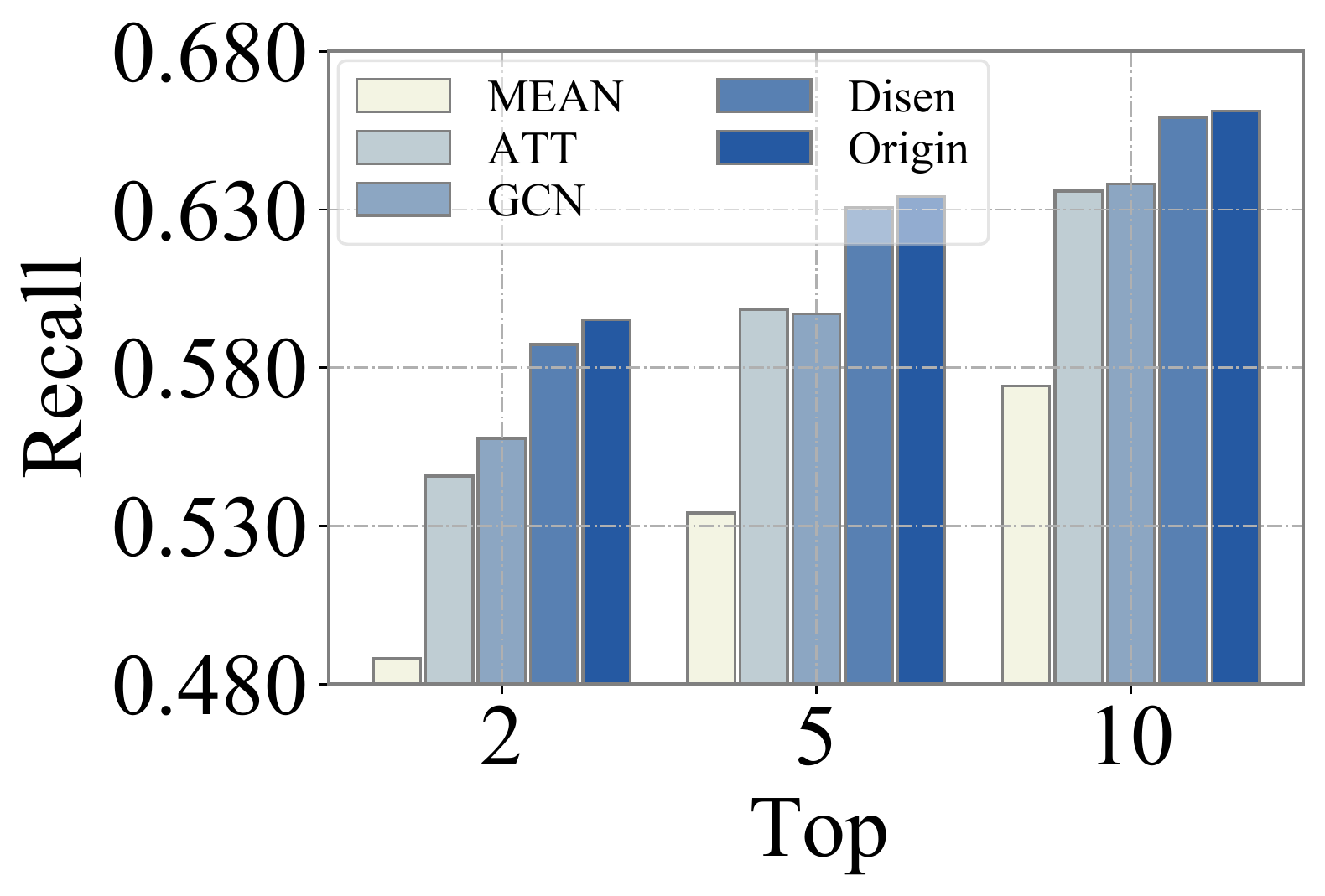}
\caption{Results on \textbf{Tokyo}}
\end{subfigure}
\begin{subfigure}{0.4\linewidth}
    \includegraphics[width=\linewidth]{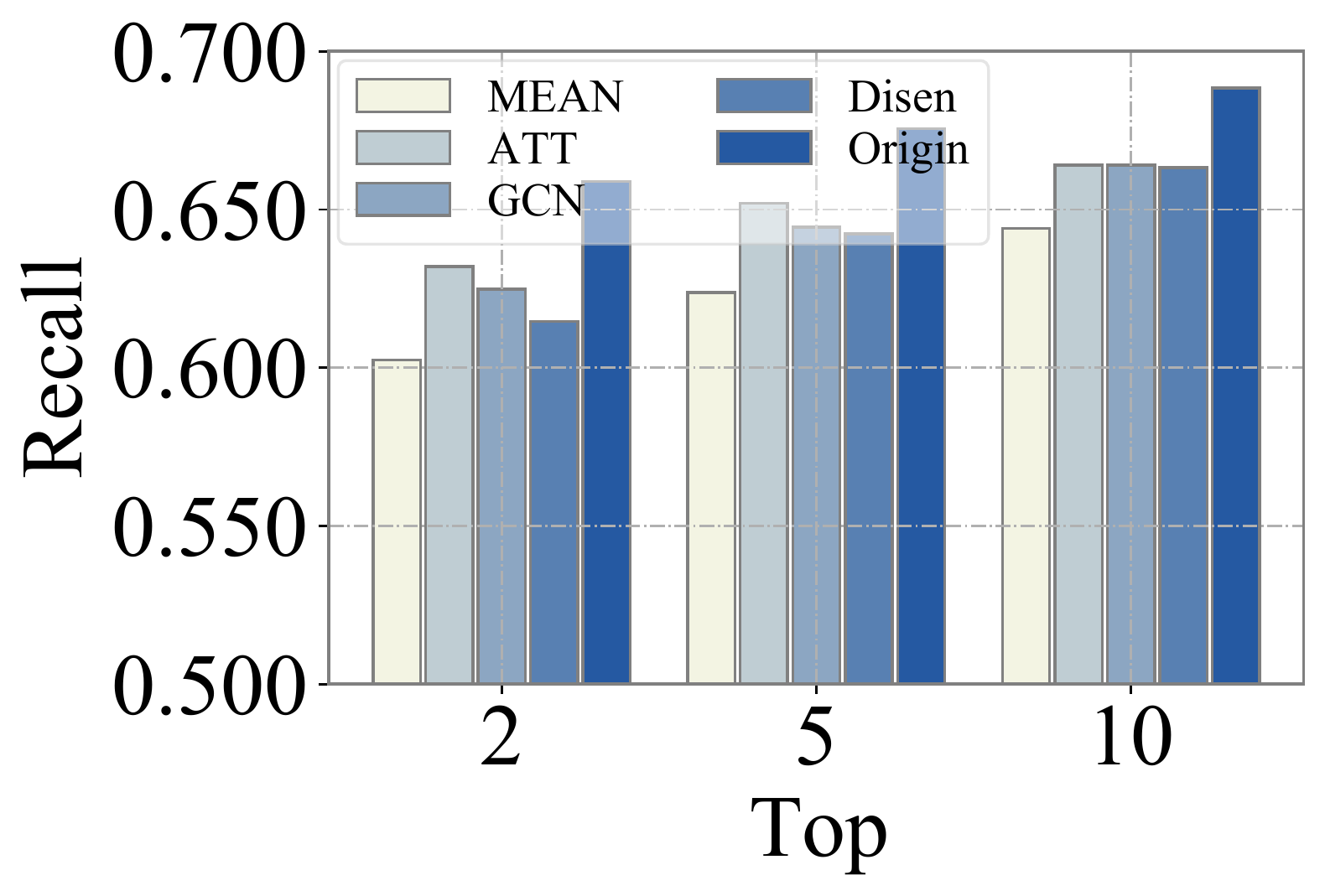}
\caption{Results on \textbf{New York City}}
\end{subfigure}
\caption{\R{Performance comparison w.r.t. different types of sequence graph encoder.}}
\label{fig:abla_seq_graph}
\end{figure}

As we have proposed to design an attention-based graph encoder to encode the spatio-temporal signals from sequence graphs, it's necessary to explore whether the proposed attention module and the message function help to obtain fine-grained user embeddings. Specifically, we compare the performance of the following variants of \method{}:

\begin{itemize} 
    \item \R{\method{}\textsubscript{Disen}: It replaces the transition graph encoder with the disentangled graph representation layer in DRAN \cite{wang2022learning}.}
    \item \method{}\textsubscript{GCN}: It replaces the transition graph encoder with a plain Graph Convolution Network (GCN) that neglects the spatio-temporal intervals of the edges.
    \item \method{}\textsubscript{ATT}: It replaces the transition graph encoder with a self-attention layer. Although this variant can aggregate information from all visited POIs, it loses the transition relationship and high-order similarities between POIs.
    \item \method{}\textsubscript{MEAN}: It drops the graph encoder and yields the user embedding merely with a mean pooling function. This variant suffers from neglecting most of the useful information.
\end{itemize}

We conduct ablation studies on the aforementioned four datasets as well. From the results illustrated in Figure \ref{fig:abla_seq_graph}, we can observe that:

\begin{itemize} 
    \item The tailor-designed graph encoder module is necessary to fully exploit the spatio-temporal information in transition graphs. The model performance suffers from information loss when the graph encoder is replaced with other structures that neglect the spatiol-temporal factors. This proves the effectiveness of our idea to design an attention-based transition graph encoder to leverage spatio-temporal intervals.
    \item The attention mechanism plays an essential role in generating fine-grained user embeddings. \method{} suffers from a significant decline when the spatio-temporal aware attention layer is removed from the model \R{(i.e., \method{}\textsubscript{Disen}, \method{}\textsubscript{GCN} and \method{}\textsubscript{MEAN})}.
    \item \R{Disentangling representations of POIs is helpful for learn more expressive POI encodings, compared with vanilla GCNs. However, the disentangling method suffers from the lack of explicit spatial and temporal information, therefore there is still a performance gap from the original graph encoder.}
    \item The neighborhood aggregation boosts the effect of the generated user embedding $x_u$, since \method{}\textsubscript{GCN} outperforms \method{}\textsubscript{MEAN}. The successive visiting pattern reveals the sequential similarities between POIs, which is helpful in modeling a user. Thus the optimal choice is to combine the spatio-temporal-based attention module with the GNN-based aggregation module to capture useful information from visiting sequences.
\end{itemize}

\subsubsection{Functionality of Spatial Preference Sampling}

\begin{table*}
\centering
\caption{The test results of \method{} and its four variants, which are \R{W/O-G.}, W/O-L., W/O-S. and W/O-C., shorts for \method{} W/O-Graph, \method{} W/O-Location, \method{} W/O-Sampling and \method{} W/O-Condition respectively. Ori. represents the original performance of \method{}. \R{\textbf{$\star$} indicates that the corresponding method outperforms the second best at the significance level with p-value<0.05.}}
\label{tab:abla_sample}
\renewcommand{\arraystretch}{1.1}
\resizebox{\linewidth}{!}{
\begin{tabular}{c ccccc ccccc}
\toprule 
\multirow{2}{*}{Model} & \multicolumn{5}{c}{Gowalla} & \multicolumn{5}{c}{Singapore} \\ 
\cmidrule[0.5pt](lr){2-6}\cmidrule[0.5pt](lr){7-11}
& \R{W/O-G.} & W/O-L. & W/O-S. & W/O-C. & Ori. & \R{W/O-G.} & W/O-L. & W/O-S. & W/O-C. & Ori. \\
\midrule
R@2 & 0.2307 & 0.2366 & 0.2431 & 0.2536 & \textbf{0.2635}\textsuperscript{\textbf{$\star$}} & 0.3019& 0.3139 & 0.3284 & 0.3408 & \textbf{0.3455}\textsuperscript{\textbf{$\star$}} \\
R@5 & 0.2774 & 0.2764 & 0.2887 & 0.3007 & \textbf{0.3086}\textsuperscript{\textbf{$\star$}} & 0.3572 & 0.3601 & 0.3715 & 0.3891 & \textbf{0.3906} \\
R@10 & 0.3177 & 0.3025 & 0.3199 & 0.3402 & \textbf{0.3421} & 0.4032 & 0.3921 & 0.4026 & 0.4258 & \textbf{0.4294}\textsuperscript{\textbf{$\star$}} \\
\midrule
N@2 & 0.2194 & 0.2177 & 0.2374 & 0.2508 & \textbf{0.2527}\textsuperscript{\textbf{$\star$}} & 0.2866 & 0.2787 & 0.3154 & 0.3301 & \textbf{0.3316}\textsuperscript{\textbf{$\star$}} \\
N@5 & 0.2389 & 0.2374 & 0.2536 & 0.2688 & \textbf{0.2716}\textsuperscript{\textbf{$\star$}} & 0.3114 & 0.3012 & 0.3271 & 0.3513 & \textbf{0.3518} \\
N@10 & 0.2561 & 0.2565 & 0.2658 & 0.2794 & \textbf{0.2872}\textsuperscript{\textbf{$\star$}} & 0.3263 & 0.3241 & 0.3396 & 0.3568 & \textbf{0.3643}\textsuperscript{\textbf{$\star$}} \\
\bottomrule 
\end{tabular}
}

\resizebox{\linewidth}{!}{
\begin{tabular}{c ccccc ccccc}
\toprule 
\multirow{2}{*}{Model} & \multicolumn{5}{c}{Tokyo}  & \multicolumn{5}{c}{New York City} \\ 
\cmidrule[0.5pt](lr){2-6}\cmidrule[0.5pt](lr){7-11}
 & \R{W/O-G.} & W/O-L. & W/O-S. & W/O-C. & Ori. & \R{W/O-G.} & W/O-L. & W/O-S. & W/O-C. & Ori. \\
\midrule
R@2 & 0.5662 & 0.5631 & 0.5894 & 0.5925 & \textbf{0.6031}\textsuperscript{\textbf{$\star$}} & 0.6000 & 0.6258 & 0.6317 & 0.6496 & \textbf{0.6589}\textsuperscript{\textbf{$\star$}} \\
R@5 & 0.6140 & 0.6047 & 0.6312 & {0.6325} & \textbf{0.6401}\textsuperscript{\textbf{$\star$}} & 0.6311 & 0.6361 & 0.6569 & {0.6751} & \textbf{0.6755}\textsuperscript{\textbf{$\star$}} \\
R@10 & 0.6480 & 0.6365 & 0.6581 & 0.6603 & \textbf{0.6681}\textsuperscript{\textbf{$\star$}} & 0.6527 & 0.6509 & 0.6695 & {0.6862} & \textbf{0.6884}\textsuperscript{\textbf{$\star$}} \\
\midrule
N@2 & 0.5503 & 0.5483 & 0.5771 & 0.5820 & \textbf{0.5886}\textsuperscript{\textbf{$\star$}} & 0.5895 & 0.6080 & 0.6204 & {0.6512} & \textbf{0.6536}\textsuperscript{\textbf{$\star$}} \\
N@5 & 0.5718 & 0.5687 & 0.5934 & {0.6021} & \textbf{0.6062}\textsuperscript{\textbf{$\star$}} & 0.6035 & 0.6273 & 0.6406 & {0.6599} & \textbf{0.6616} \\
N@10 & 0.5829 & 0.5784 & 0.6027 & {0.6122} & \textbf{0.6160}\textsuperscript{\textbf{$\star$}} & 0.6104 & 0.6255 & 0.6575 & {0.6706} & \textbf{0.6745}\textsuperscript{\textbf{$\star$}} \\
\bottomrule 
\end{tabular}
}
\end{table*}

The diffusion-based sampling strategy is intended for sampling from the user-specific posterior distribution. We expect the model to be capable of modeling different locational preferences and generating more instructive representations for recommending a next-to-visit POI. Thus it is necessary to figure out whether the sampling module improves the recommendation results. Specifically, we consider four variants of \method{}, \R{which are: \method{} without all graph encoders, \method{} without the user-specific context condition and the score network is replaced with an MLP with just $\mathrm{v}_u(t)$ as input, \method{} without the entire sampling module where the prototype $\hat{\mathrm{v}}_u$ is directly fed into the calculation of recommendation score, and \method{} without any locational embeddings at all. The aforementioned variants are named as W/O-Graph, W/O-Condition, W/O-Sampling, and W/O-Location respectively. To ensure robustness and reliability of our results, we conduct each setting five times with different random seeds and report the average performance across these iterations.} The comparison result between the original \method{} and the variants is reported in Table \ref{tab:abla_sample} and we can observe that:

\begin{itemize} 
    \item \R{The sequential and geographical graph encoders are necessary for \method{} to extract informative POI embeddings for downstream recommendation.} \R{The extra locational embedding plays a vital role in location-based POI recommendations for models without any locational embeddings performs worse than most of the other variants.} Explicitly representing the geographical information brings significant improvements to model performance. The model is able to assess whether the target item is similar to the user's locational preference by the encodings of the distance-based POI graph.
    \item The diffusion-based sampling process is necessary to improve the recommendation results for the model without the sampling process under-performs the original model and the model without conditional sampling. The sampling strategy brings over 3.3\% and 2.5\% improvement on Recall@10 and NDCG@10 respectively, which proves the effectiveness of the sampling module. The diffusion-based sampling process could generate a more relevant and user-specific locational preference and boost the recommendation process.
    \item The user-specific context condition can bring positive influences for sampling an effective locational preference. Though compared with the model without sampling, \method{} could still benefit directly from the locational posterior, yet it suffers a performance decline for neglecting the user-specific information provided during the controlled diffusion process. 
\end{itemize}

\subsubsection{Effect of Fisher Loss}

\begin{figure}[t]
\centering

\begin{subfigure}{0.4\linewidth}
    \includegraphics[width=\linewidth]{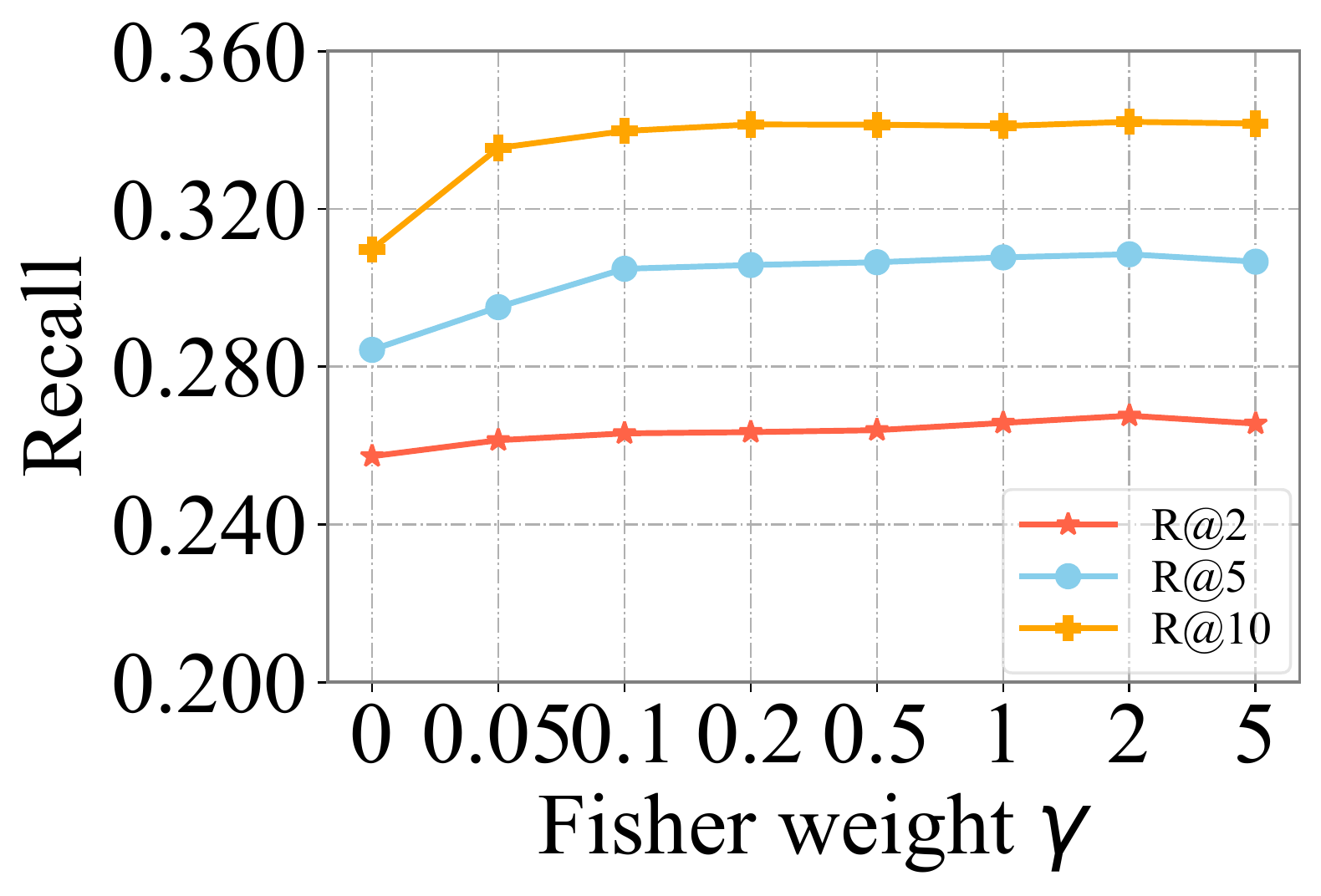}
\caption{Results on \textbf{Gowalla}}
\end{subfigure}
\begin{subfigure}{0.4\linewidth}
    \includegraphics[width=\linewidth]{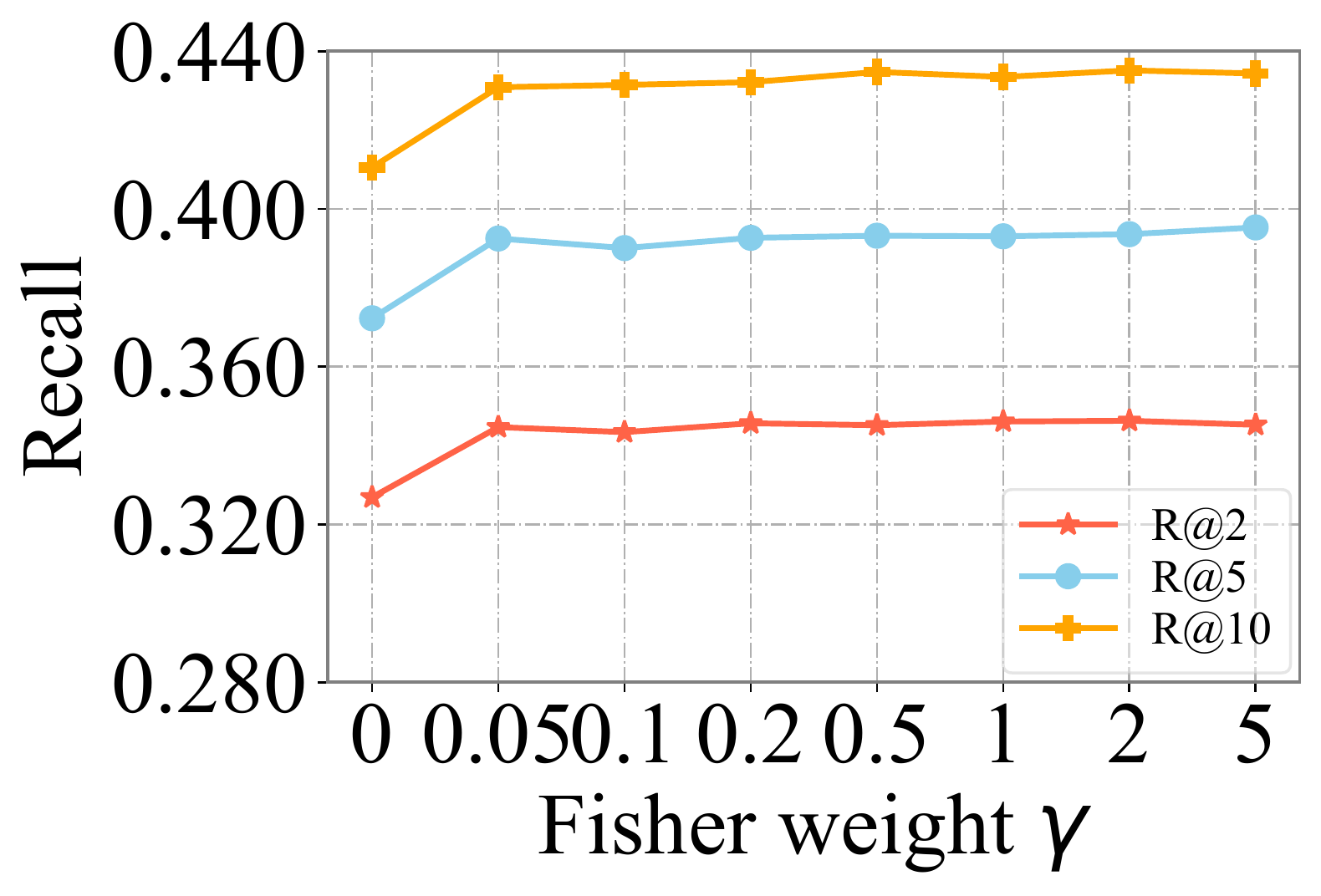}
\caption{Results on \textbf{Singapore}}
\end{subfigure}


\begin{subfigure}{0.4\linewidth}
    \includegraphics[width=\linewidth]{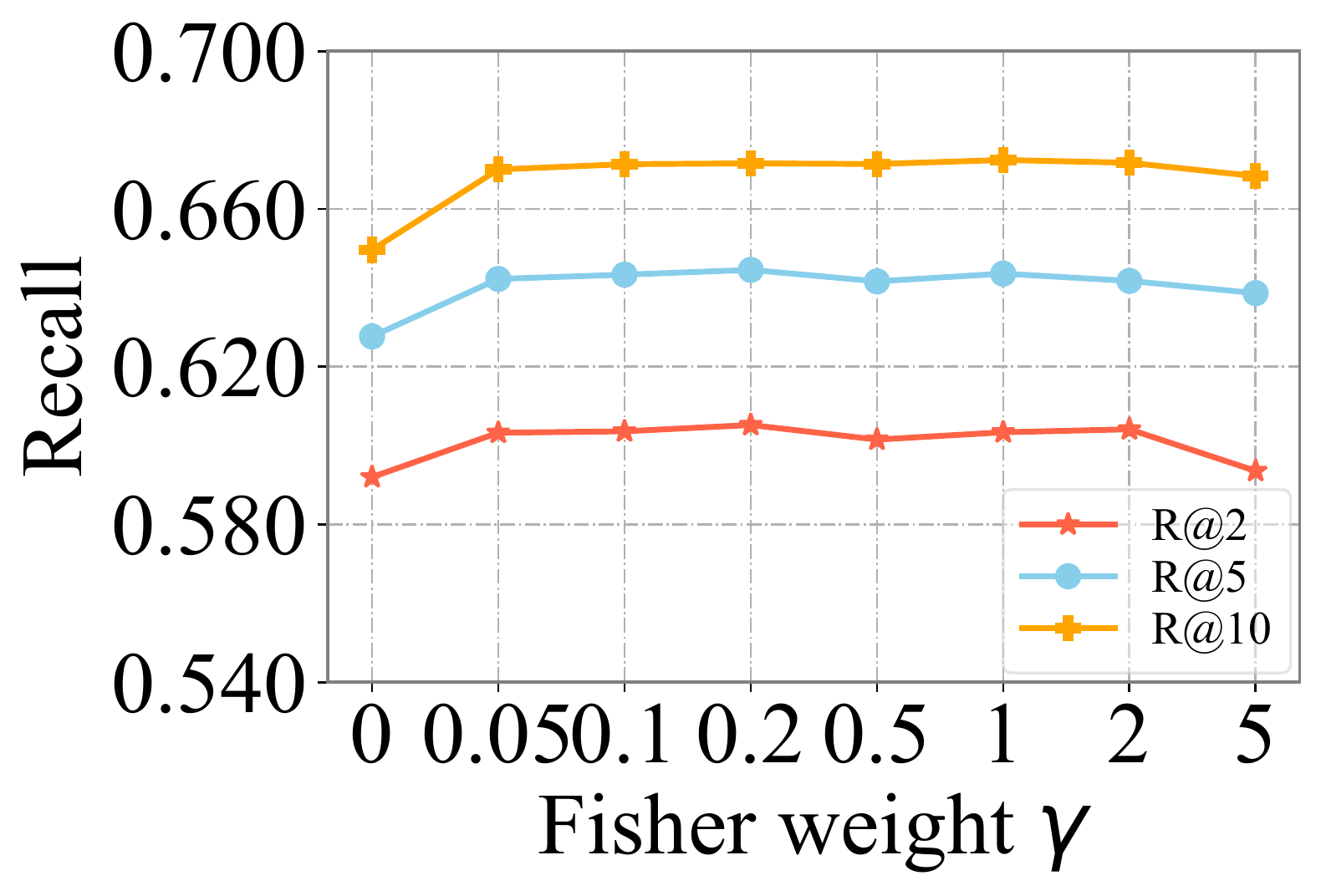}
\caption{Results on \textbf{Tokyo}}
\end{subfigure}
\begin{subfigure}{0.4\linewidth}
    \includegraphics[width=\linewidth]{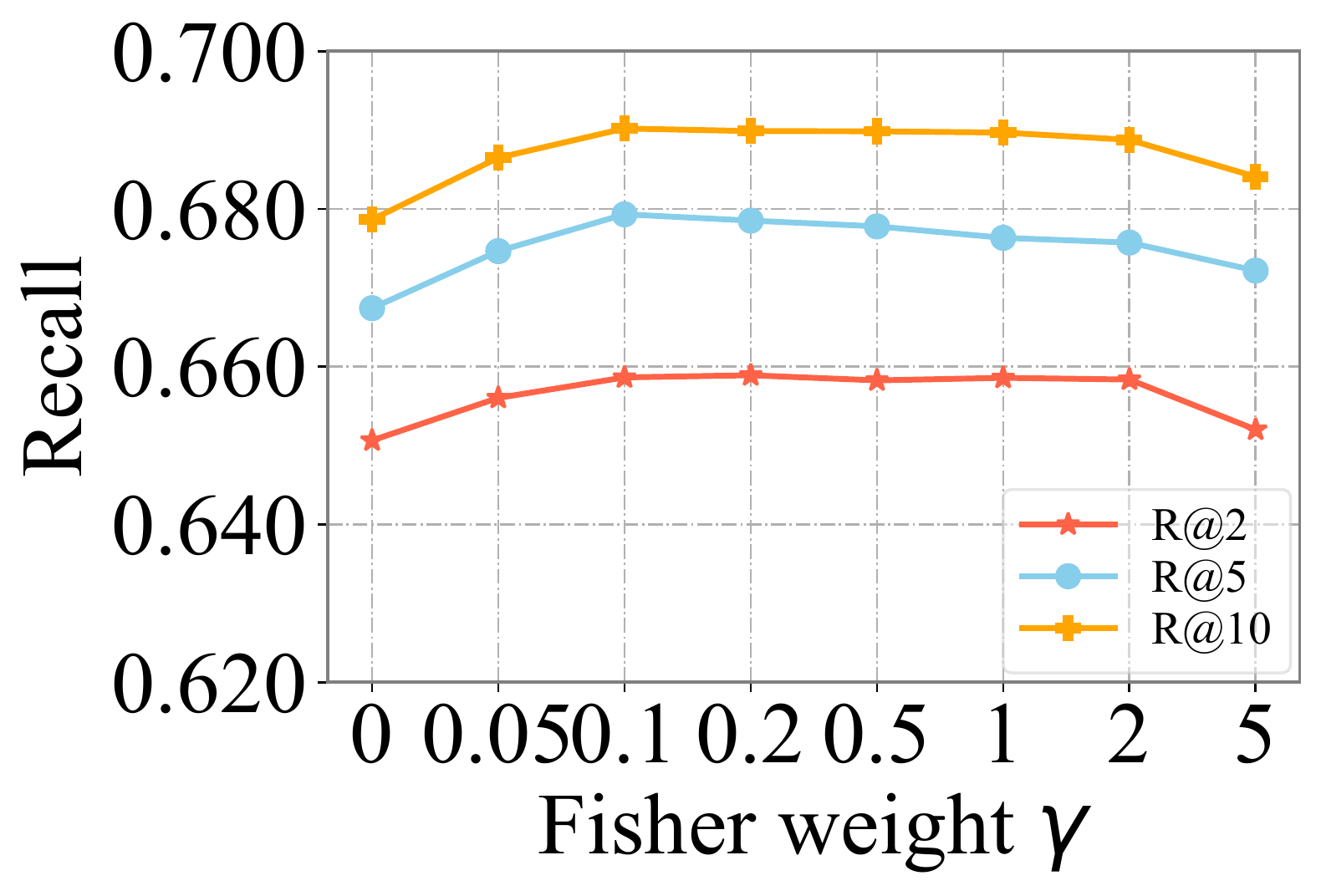}
\caption{Results on \textbf{New York City}}
\end{subfigure}
\caption{Recommendation recall w.r.t. different weight coefficient $\gamma$ on four datasets.}
\label{fig:param_fisher_weight}
\end{figure}

Recall that we adopt a Fisher loss defined by Eq. \ref{eq:fisher_loss} as the optimization target in Eq. \ref{eq:total_loss}, where the loss is controlled by a hyper-parameter $\gamma$. The Fisher loss $\mathcal{L}_{Fisher}$ is proposed to optimize the score function $s_\theta$ so the model can sample a more accurate location preference that is close to the target embedding. We wonder how $\mathcal{L}_{Fisher}$ affects the recommendation performance, so we conduct a parameter study on $\gamma$. Especially, we vary $\gamma$ from 0 (without $\mathcal{L}_{Fisher}$) to 5 and record the model performance. From the results in Figure \ref{fig:param_fisher_weight} we can observe that:

\begin{itemize} 
    \item The Fisher loss $\mathcal{L}_{Fisher}$ is necessary for the promising model performance and $\mathcal{L}_{Fisher}$ regularizes the sampling module to more accurate generation results. When $\gamma$ is 0 and no regularization is applied on the score network, the model performance declines due to the degenerated score function and fails to sample a representative locational preference and influence the model performance.
    \item The model performance reaches the optimal when $\gamma$ is around 0.2 to 1. When the regularization is weak, the model lacks a reasonable sampling module and fails to model the locational preference. However on some datasets, when $\gamma$ is extremely large (over 2.0), \method{} would overly focus on generating an accurate locational embedding and ignore the useful information from the user's visiting history.
    \item Even though the model performance suffers when $\gamma$ is reduced to 0, \method{} can still achieve competitive performance against most of the baseline methods. This implies that the score network $s_\theta(\cdot)$ can be optimized with the main cross entropy loss $\mathcal{L}_{CE}$. As the auxiliary fisher loss $\mathcal{L}_{Fisher}$ boosts the optimization of $s_\theta(\cdot)$ with additional assumptions on the marginal probability throughout the diffusion process, it can bring extra benefits to model performance.
\end{itemize}

\subsubsection{Effect of Sampling Steps}

\begin{figure}[t]
\centering

\begin{subfigure}{0.4\linewidth}
    \includegraphics[width=\linewidth]{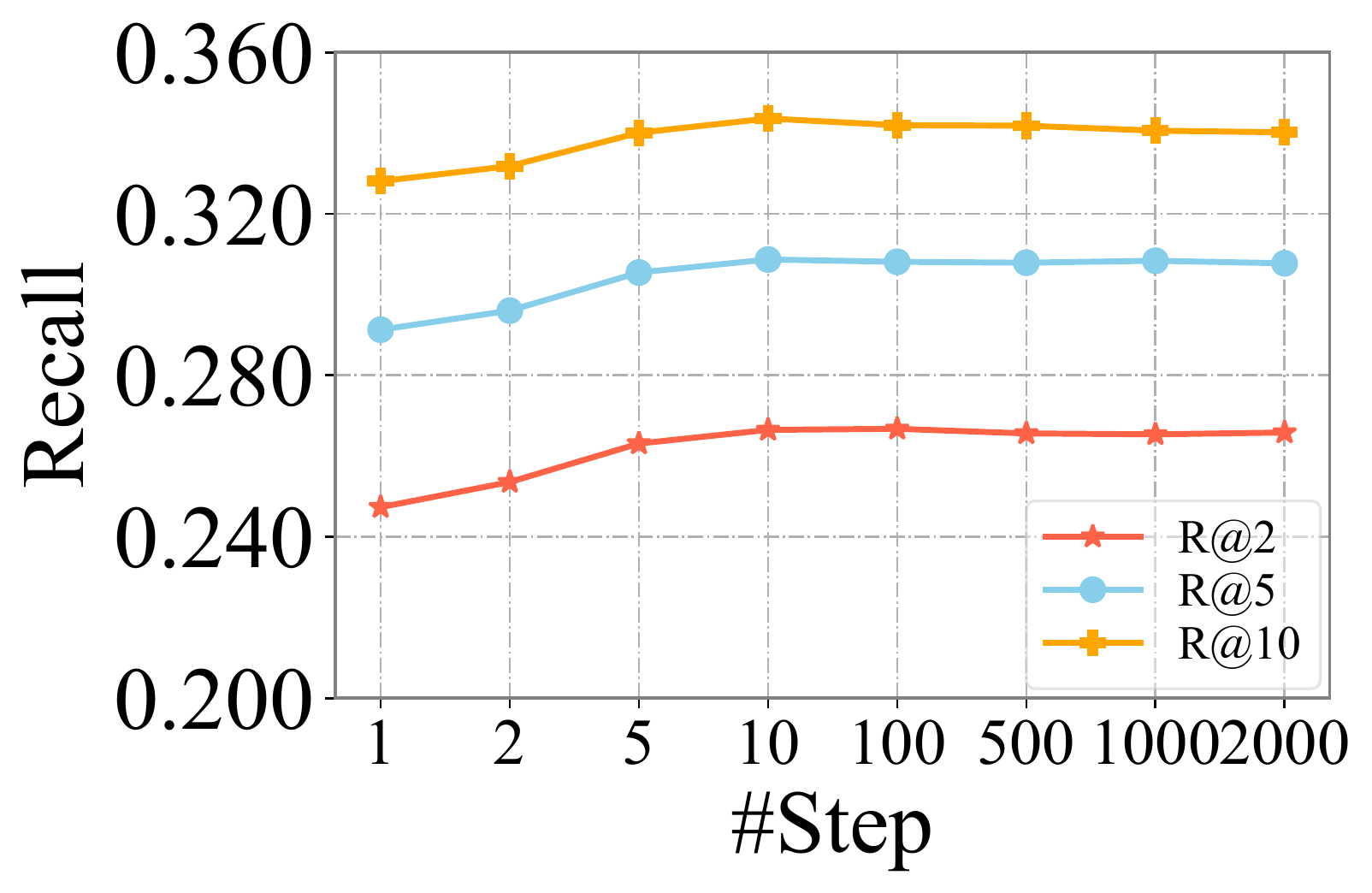}
\caption{Results on \textbf{Gowalla}}
\end{subfigure}
\begin{subfigure}{0.4\linewidth}
    \includegraphics[width=\linewidth]{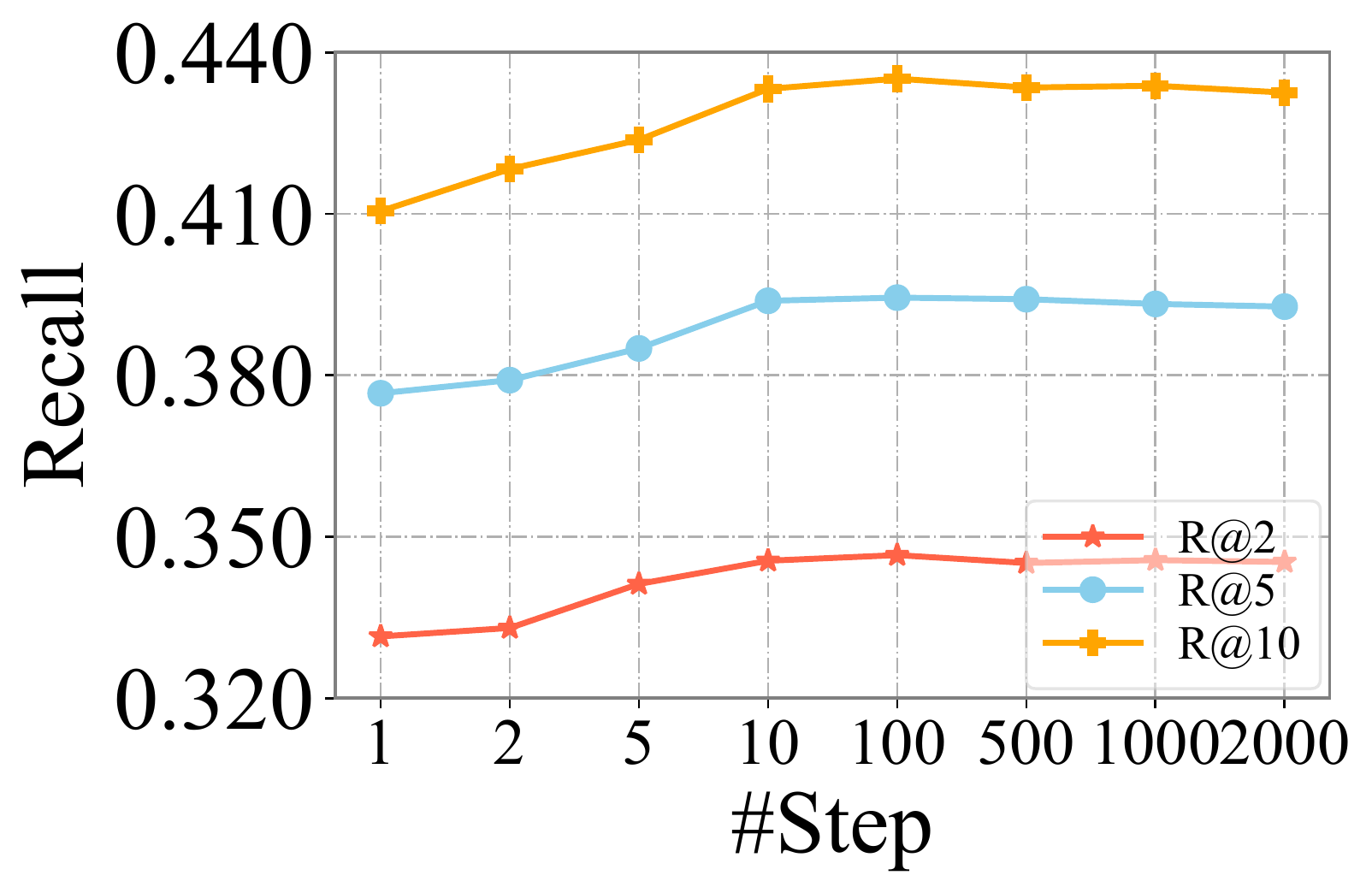}
\caption{Results on \textbf{Singapore}}
\end{subfigure}


\begin{subfigure}{0.4\linewidth}
    \includegraphics[width=\linewidth]{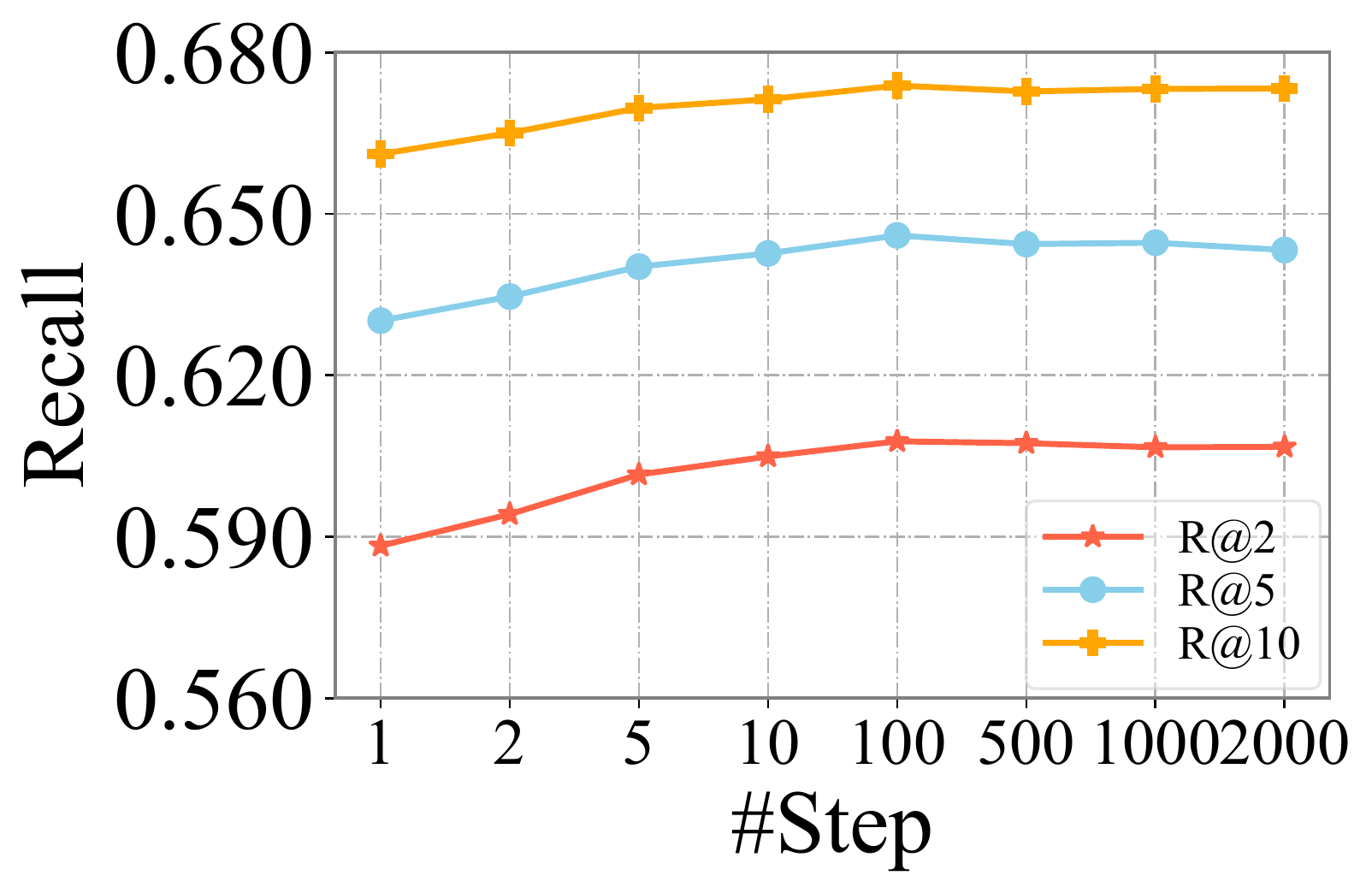}
\caption{Results on \textbf{Tokyo}}
\end{subfigure}
\begin{subfigure}{0.4\linewidth}
    \includegraphics[width=\linewidth]{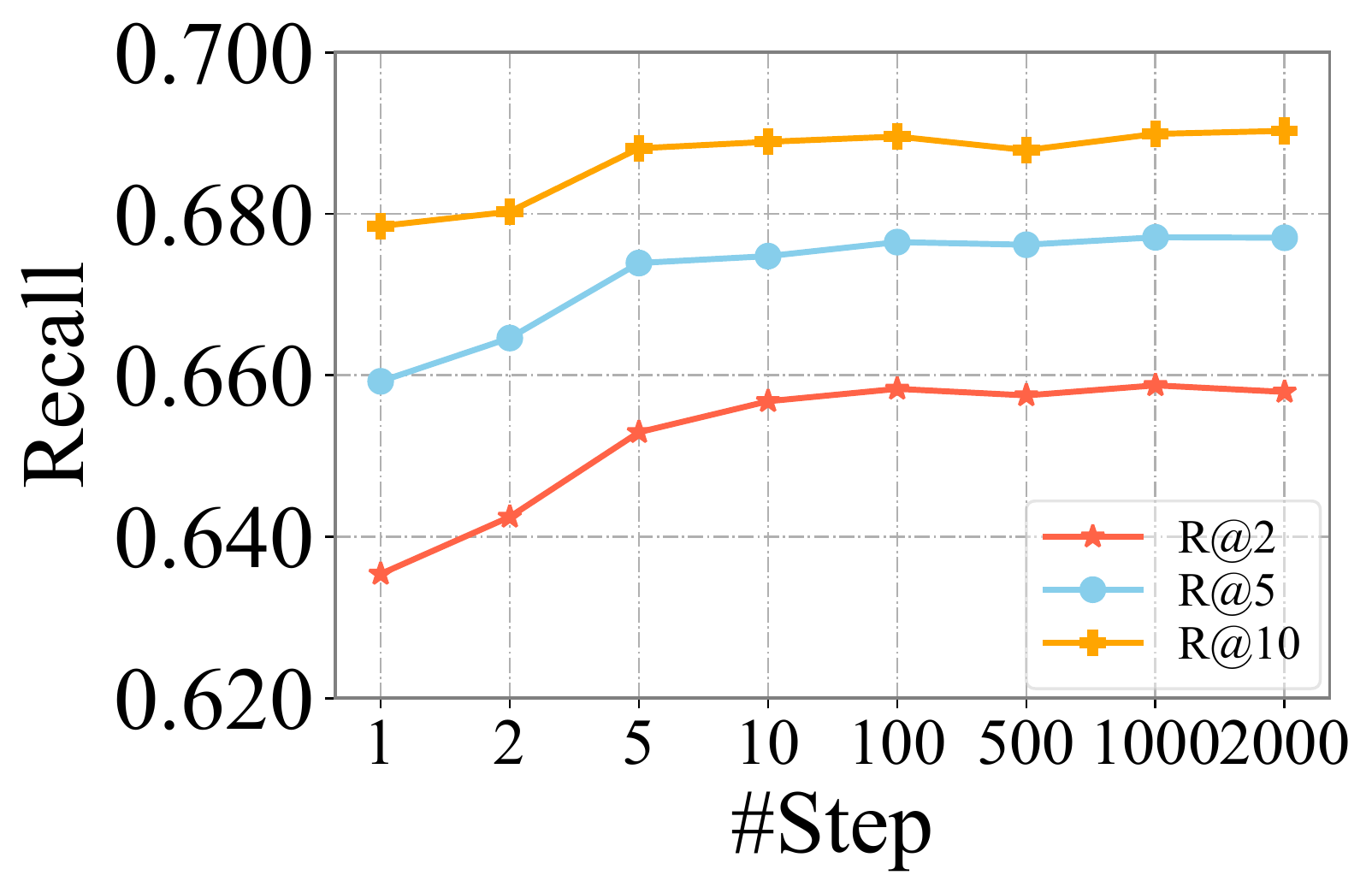}
\caption{Results on \textbf{New York City}}
\end{subfigure}
\caption{Recommendation recall w.r.t. different numbers sampling steps.}
\label{fig:param_sde_step}
\end{figure}
Recall that for the sampling process, we adopt a fixed step size $\mathrm{d}t$ for solving the reversed SDE defined in Eq. \ref{eq:sde_backward}. Intuitively, with little step size, the model is capable of capturing fine-grained information for being better fitted to the assumption of $\Delta t\rightarrow0$ in diffusion process \cite{ho2020denoising}. However, the specific influence of the step size $\mathrm{d}t$ remains to be explored. We explore the performance of \method{} by varying $\mathrm{d}t$ from 1 (where the sampling process degenerates to a single-step adjustment) to $10^{-4}$ and the results are reported in Figure \ref{fig:param_sde_step}. We make the following observations:

\begin{itemize} 
    \item Generally $\mathrm{d}t$s close to 0 can lead to better results since the model performance increases when the step size goes smaller $\mathrm{d}t$. By decreasing the step size, we actually extend the step number of the reversed diffusion process and empower the sampling module by capturing more fine-grained gradient information.
    \item Even the one-step sampling module ($\mathrm{d}t=1$) outperforms the model without sampling process. The result indicates that the model could still learn more about the user-specific locational preference. However, it doesn't hold for the assumption about $\mathrm{d}t\rightarrow 0$ and suffers from the performance decline.
    \item When step size is below a specific value (e.g. 100), the performance no obvious increase in model performance can be observed. This result makes the choosing of $\mathrm{d}t$ a trade-off between model performance and computational consumption since increasing sampling steps will significantly increase the time complexity of \method{}.
\end{itemize}

\subsection{In-depth Study (RQ3)}
Recall that we propose \method{} to better clarify the user's spatial preference in a more fine-grained manner. Specifically, the diffusion-based sampling module solves a reverse diffusion SDE to dynamically sample from the user's preference posterior. We wonder if the proposed sampling strategy helps to depict the user's spatial preference effectively. Thus we expect to intuitively show how the sampling result approaches the true locational preference of the user. \R{In order to evaluate the effectiveness of \method{} in addressing different user preferences, we divide users into distinct groups based on their geographical preferences , and compare \method{}'s performance with different baseline methods for each user group.}
\R{Additionally, we select three representative users, namely user A, B and C. User A represents users who tend to follow regular routines and visit the same POIs frequently. User B frequently visits two different neighborhoods, indicating a preference for diverse locations. User C is an adventurous user who seeks out new and unfamiliar areas. We visualize the top-recommended POIs with the sampled location preference at different sampling steps for these users. Furthermore, we conduct detailed case studies to examine whether the sampling strategy is capable of recommending POIs with personalized locational preferences.}

\subsubsection{Model Performance on Different User Groups}
\begin{figure}[t]
\centering

\begin{subfigure}{0.4\linewidth}
    \includegraphics[width=\linewidth]{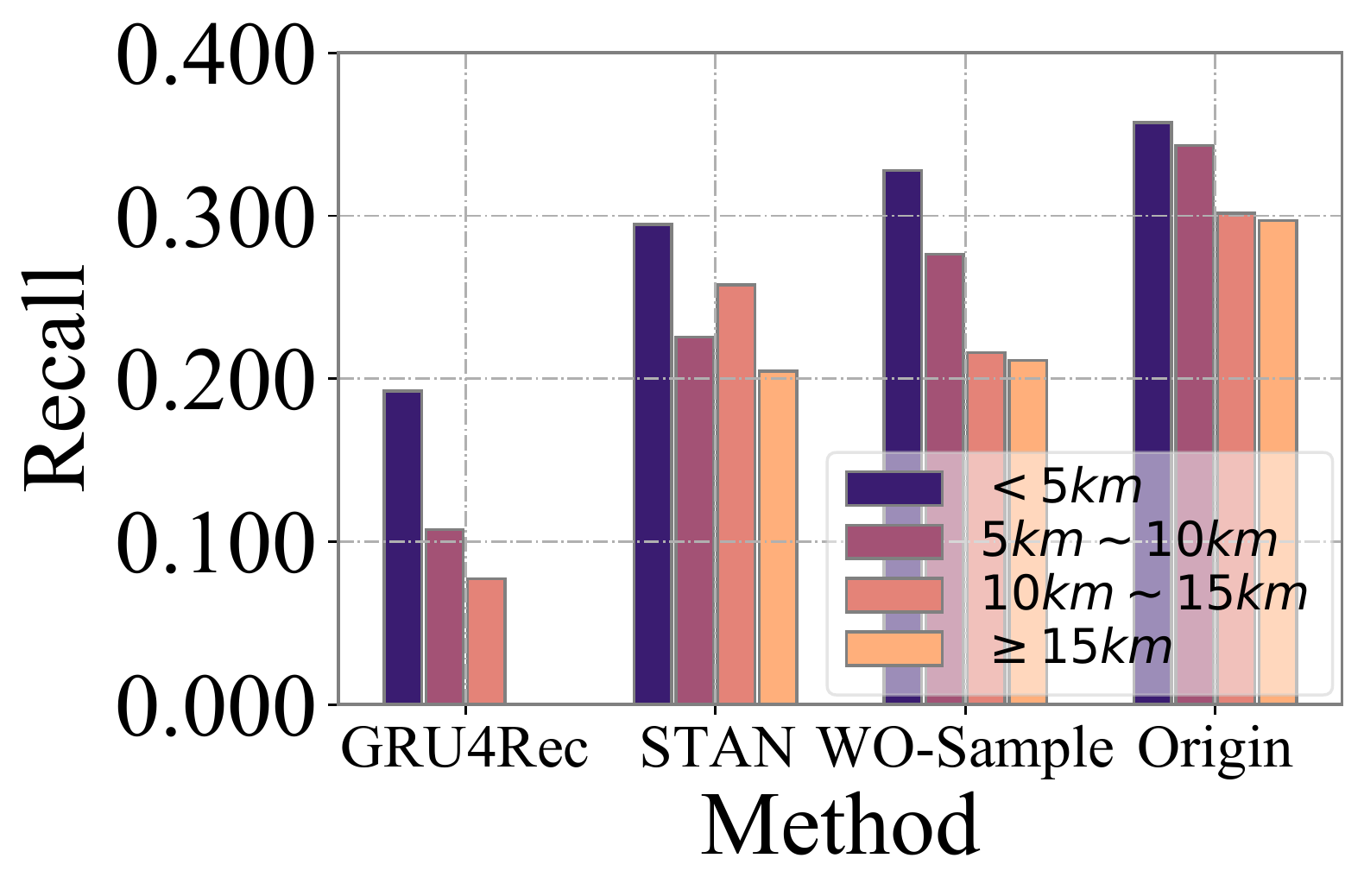}
\caption{Results on \textbf{Singapore}}
\end{subfigure}
\begin{subfigure}{0.4\linewidth}
    \includegraphics[width=\linewidth]{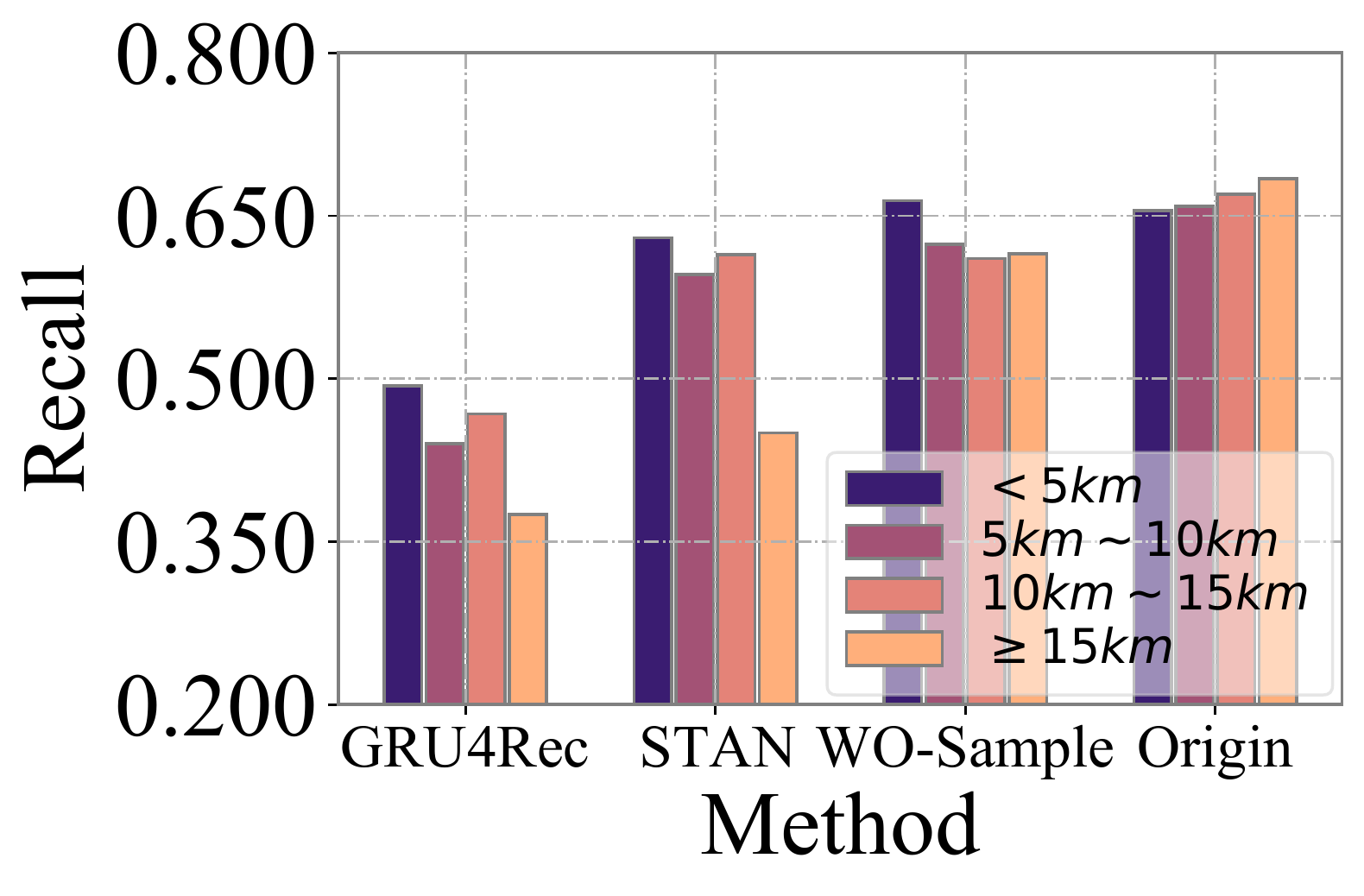}
\caption{Results on \textbf{New York City}}
\end{subfigure}

\caption{\R{Recommendation recall w.r.t. different user groups}}
\label{fig:user_group}
\end{figure}

\R{To assess whether the proposed sampling strategy of \method{} effectively captures different locational preferences, we divide users into different groups based on their geographical visiting patterns and evaluate the model's performance for each group. Specifically, we utilize datasets from Singapore and New York City and calculate the average distance $\bar{\delta_s}$ between successive visits for each user. Larger values of $\bar{\delta_s}$ indicate more adventurous users who travel longer distances between visits. We categorize users into different groups based on their average distance between successive visits: from visitors in a limited area ($\bar{\delta_s} < 5$ km) to long-distance travelers ($\bar{\delta_s} \geq 15$ km). For each group, we measure the Recall@2 performance of four methods: GRU4Rec, STAN, \method{} without the sampling module, and the original \method{}. The results are presented in Figure \ref{fig:user_group}:}

\R{
\begin{itemize}
    \item In general, as the average distance $\bar{\delta_s}$ between successive visits increases, it becomes more challenging for a model to accurately recommend locations for users. This is because the locational preferences of active users who travel longer distances are harder to capture and model effectively.
    \item Methods that explicitly incorporate the locational influence in historical visits (STAN and \method{}) show improvement compared to traditional methods and mitigate the decline in performance. Notably, the proposed \method{} consistently outperforms other methods and maintains robustness even as the average distance increases. This highlights the importance and effectiveness of the diffusion-based sampling module in \method{}. For some specific cases (e.g. users who are less adventurous in New York City dataset), \method{} without sampling module may achieve good performance for successfully depicting the local visiting structures.
\end{itemize}
}

\subsubsection{\R{Visualization of Dynamic Sampling}}

\begin{figure}[t]
\centering
\begin{subfigure}{\linewidth}
\includegraphics[width=\linewidth]{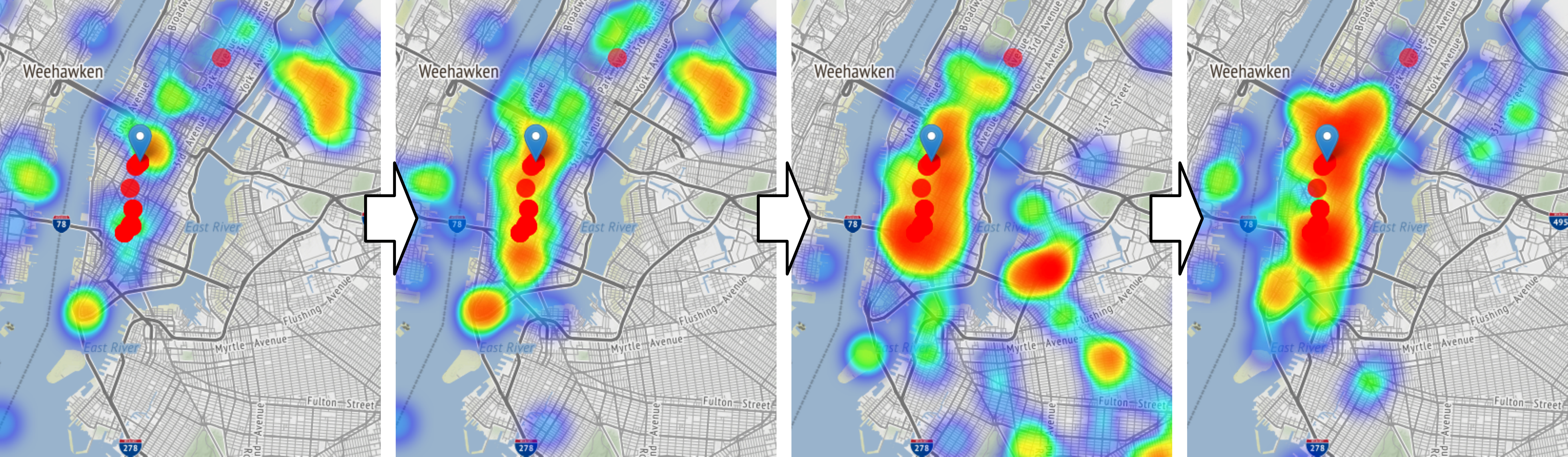}
\caption{Experiment results for user A.}
\label{fig:case_heatmap_A}
\end{subfigure}

\begin{subfigure}{\linewidth}
\includegraphics[width=\linewidth]{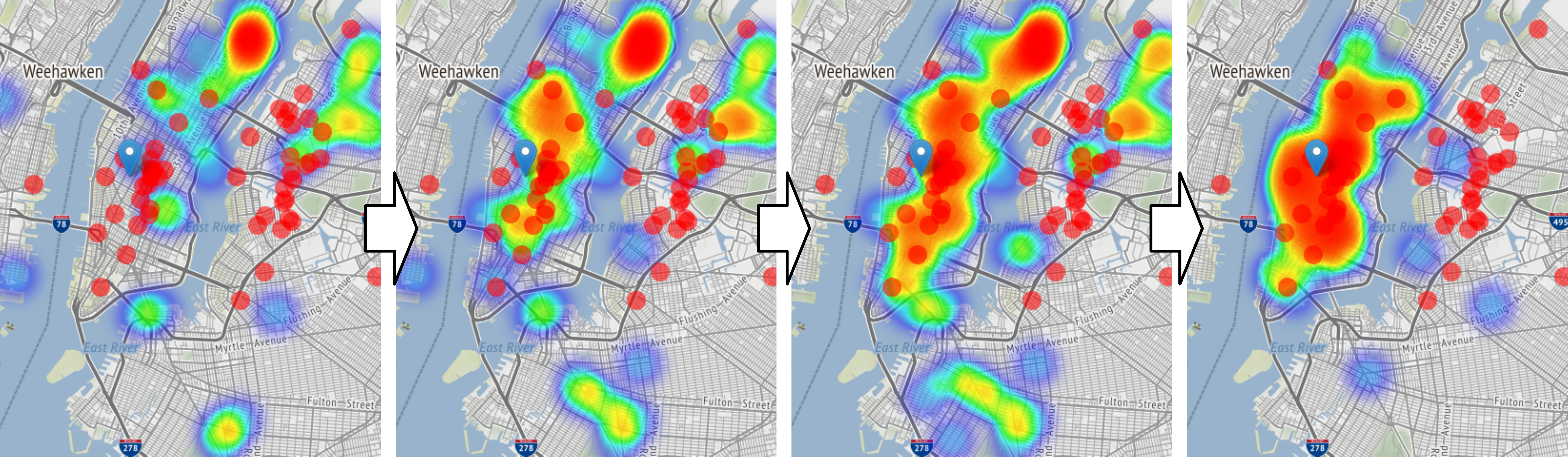}
\caption{Experiment results for user B.}
\label{fig:case_heatmap_B}
\end{subfigure}

\begin{subfigure}{\linewidth}
\includegraphics[width=\linewidth]{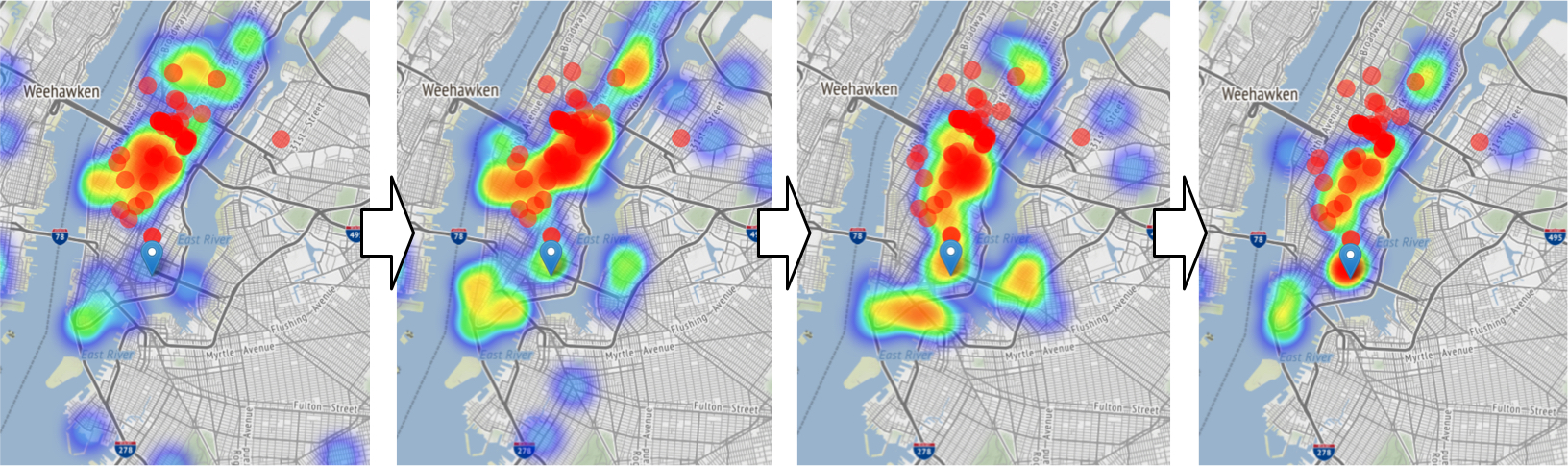}
\caption{\R{Experiment results for user C.}}
\label{fig:case_heatmap_C}
\end{subfigure}
\caption{Visualization of the top-100 recommended POIs as the diffusion-based sampling process proceeds. Each user's previously visited POIs are marked with red dots and the target POI is marked with blue markers.}
\label{fig:case_heatmap}
\end{figure}

The diffusion-based sampling module samples the locational preference in an iterative manner, which means that the sampled data should eventually approach the real distribution, in our case is the location of the target POI. \R{To visualize this process, we train \method{} on New York City dataset, and obtain the sampled locational preference $\mathrm{v}_u$ for users A, B and C when the sampling process is 0\% (the initial value), 33\%, 67\%, and 100\% complete. We then calculate the top-100 recommended POIs with the sampled $\mathrm{v}_u$ only and visualize them in heat maps.} The history POIs are marked with red dots and the targets are marked with blue markers. From the the results in Figure \ref{fig:case_heatmap} we can observe that:

\begin{itemize}
    \item With the help of the sampling strategy, \method{} is capable of finding the approximate location for the next visits. From both figures, we can observe a clear pattern that the recommended POIs are clustering around the location of the ground truth target POI as the sampling process proceeds. This intuitively shows the diffusion-based sampling process is helpful to figure out the locational preference of specific users.
    \item Predicting POIs for regular users like A is easier for the sampling module to depict. As illustrated in Figure \ref{fig:case_heatmap_A}, the historical visits of A show clear regional characteristics (basically around the Manhattan district), making it easy to locate the possible POIs around the familiar areas at the early stage of the sampling process. In contrast, the visited POIs of user B are scattered around the city (in Manhattan, Queens and Brooklyn) so it takes more steps to sample the true preference of B and locate the target POI through the sampling process.
    \item \R{For adventurous users like User C. The sampling module plays a crucial role in capturing their future locational preferences. We can observe that even when the target POI is far from User C's historical visits, the sampling process of \method{} gradually guides the model to recommend POIs in the neighborhood around the ground truth. The observation reflects the capability of the sampling module in exploring personalized locational preference.}
    \item Though the sampled locational preference can intuitively reflect which region the user is most likely to visit, it is unable to directly find the accurate location of target POI. As illustrated, when the sampled process converges, the predicted POIs are gathering around a relatively large area around target POI. This implies that we still need a sequence-based preference score to accurately select the target POI from a wide range of possible candidates.
\end{itemize}

\subsubsection{Case Study of Sampling Process}

\begin{table*}
\centering
\caption{The top-5 recommended POIs via the sampled spatial preference of users. The category and distance (in km) between the recommended POI and target POI (i.e. ground truth) is recorded in the table. The category marked with $\star$ represents the target POI.}
\label{tab:case_sample}
\renewcommand{\arraystretch}{1.1}
\resizebox{\linewidth}{!}{
\begin{tabular}{c cc cc cc cc}
\toprule 
\multicolumn{9}{c}{{Recommended POIs for User A w.r.t. sampling step}} \\
\bottomrule 
\multirow{2}{*}{Top-K} & \multicolumn{2}{c}{0\%} & \multicolumn{2}{c}{33\%} & \multicolumn{2}{c}{66\%} & \multicolumn{2}{c}{100\%} \\ 
\cmidrule[0.5pt](lr){2-3} \cmidrule[0.5pt](lr){4-5} \cmidrule[0.5pt](lr){6-7} \cmidrule[0.5pt](lr){8-9}
& Category & Distance & Category & Distance & Category & Distance & Category & Distance \\
\midrule
1 & Seafood & 3.55 & Tattoo Parlor & 2.02 & Caribbean Food$^\star$ & 0.00 & Caribbean Food$^\star$ & 0.00 \\
2 & Mediterranean Food & 5.24 & Seafood & 3.55 & Vegetarian Food & 1.84 & Home (private) & 1.91 \\
3 & Bank & 2.39  & Bar & 3.41 & Coffee Shop & 1.41 & American Food & 0.53 \\
4 & Park & 6.76  & Caribbean Food$^\star$ & 0.00 & Sandwich Place & 2.88 & Seafood & 3.55 \\
5 & Food \& Drink & 4.57 & Clothing Store & 6.86 & Bus Station & 3.88 & Subway & 1.89 \\

\bottomrule 
\end{tabular}
}

\resizebox{\linewidth}{!}{
\begin{tabular}{c cc cc cc cc}
\toprule 
\multicolumn{9}{c}{{Recommended POIs for User B w.r.t. sampling step}} \\
\bottomrule
\multirow{2}{*}{Top-K} & \multicolumn{2}{c}{0\%} & \multicolumn{2}{c}{33\%} & \multicolumn{2}{c}{66\%} & \multicolumn{2}{c}{100\%} \\ 
\cmidrule[0.5pt](lr){2-3} \cmidrule[0.5pt](lr){4-5} \cmidrule[0.5pt](lr){6-7} \cmidrule[0.5pt](lr){8-9}
& Category & Distance & Category & Distance & Category & Distance & Category & Distance \\
\midrule
1 & Burger Joint & 7.35 &  Burger Joint & 7.35 & Outdoor Store & 1.31 & Laundry Service & 0.59\\
2 & Smoke Shop & 9.25 & College Building & 10.15 & American Food & 1.83 & Bar & 0.69 \\
3 & Food \& Drink & 15.13 & Subway & 4.07 & Bridge & 1.94 & 
 Pharmacy$^\star$ & 0.00  \\
4 & Medical Center & 15.61 & Greek Food & 3.24 & Pharmacy$^\star$ & 0.00 & Bakery & 1.33\\
5 & Bar & 9.62 & Medical Center & 15.61 & Food \& Drink & 1.03 & Deli & 0.97 \\

\bottomrule 
\end{tabular}
}

\resizebox{\linewidth}{!}{
\begin{tabular}{c cc cc cc cc}
\toprule 
\multicolumn{9}{c}{\R{Recommended POIs for User C w.r.t. sampling step}} \\
\bottomrule
\multirow{2}{*}{Top-K} & \multicolumn{2}{c}{0\%} & \multicolumn{2}{c}{33\%} & \multicolumn{2}{c}{66\%} & \multicolumn{2}{c}{100\%} \\ 
\cmidrule[0.5pt](lr){2-3} \cmidrule[0.5pt](lr){4-5} \cmidrule[0.5pt](lr){6-7} \cmidrule[0.5pt](lr){8-9}
& Category & Distance & Category & Distance & Category & Distance & Category & Distance \\
\midrule
1 & Bar & 6.59 & Apartment & 10.01 & Gastropub & 1.36 & Gastropub & 1.36\\
2 & Gastropub & 1.36 & Bar & 6.59 & Bar & 6.59 & Bar & 6.59 \\
3 & Apartment & 10.01 & Neighborhood & 4.07 & Spa \& Massage$^\star$ & 0.00 & 
Spa \& Massage$^\star$ & 0.00  \\
4 & Stadium & 5.14 & Gastropub & 1.36 & Bridge & 1.34 & Diner & 0.92\\
5 & Sushi Restaurant & 3.22 & Sushi Restaurant & 3.22 & Diner & 0.92 & Bridge & 1.34 \\

\bottomrule 
\end{tabular}
}
\end{table*}

To further investigate the effectiveness of the sampling module, we conduct a detailed case study by recording the top-5 recommended POIs for both users throughout the sampling process. Particularly, we calculate the recommendation score with the sampled locational preference when the sampling process is 0\%, 33\%, 67\%, and 100\% done respectively and recall the top-5 recommended POIs sorted by the score. We report the category of the recalled POIs and the distance from recalled POIs and corresponding target POIs. From the results in Table \ref{tab:case_sample} we can observe that:

\begin{itemize}
    \item The sampled representations reflect the locational characteristics of POIs. The recalled POIs are more likely to be in the same region, for the distance between the target and recalled POIs are similar at every step. The distances between the recalled POIs and target POIs are getting smaller when the sampling step grows, which illustrates that the sampling module is capable of locating the target POI via the reversed diffusion process.
    \item Generally recommending POIs to users with regular routines (e.g. user A) is more accurate since the helpful information from the previous visiting patterns can be easily leveraged. With the expressiveness patio-temporal sequence graph encoder, \method{} is able to recall the target POI in the early stage of the sampling process. The other recalled POIs are geographically close to the target POI, which makes them reasonable candidates as well.
    \item However, for users who are exploring novel areas (e.g. user B and C), simply aggregating the representations of historical visits would not be as effective. \R{Meanwhile, as the sampling step increases, \method{} successfully samples the user's implicit locational preference, and there are more and more nearby POIs appear in the top-5 recommended list. The target POIs are eventually found out by the model as well.}
    The result illustrates the necessity to iteratively sample the locational preference to locate the accurate spot of the next-to-visit POIs.
\end{itemize}

\section{Conclusion}
In this paper, we propose a novel graph-based POI recommendation model, namely \method{}, to exploit the spatio-temporal transition graph and model the user's locational preference. \method{} is equipped with a tailor-designed sequence graph encoder that depicts the user's spatio-temporal trajectory, followed by a diffusion-based sampling strategy to sample the locational preference of the user. Comprehensive experiments on four real-world datasets demonstrate the effectiveness of the proposed \method{} framework and its superiority over the existing state-of-the-art baseline models. Ablation and case studies further provide an intuitive view of how the sampling process depicts the user's hidden locational preference.

\begin{acks}
The authors are grateful to the anonymous reviewers for critically reading the manuscript and for giving important suggestions to improve their paper. 

This paper is partially supported by the National Natural Science Foundation of China with Grant (NSFC Grant Numbers 62306014 and 62276002) as well as the China Postdoctoral Science Foundation with Grant No. 2023M730057.
\end{acks}

\bibliographystyle{ACM-Reference-Format}
\bibliography{main}

\end{document}